\documentclass[preprint,aps,prd,nofootinbib,superscriptaddress]{revtex4}
\usepackage{hyperref}
\usepackage[pdftex]{graphicx,color}
\usepackage{slashed}
\usepackage{amsmath,amssymb}
\usepackage{hhline}
\usepackage{subfigure}
\usepackage{epstopdf}
\usepackage{calc}

\newcommand{\ket}[1]{\left|#1\right\rangle}
\newcommand{\bra}[1]{\left\langle#1\right|}

\newcommand{\beq}{\begin{equation}}
\newcommand{\eeq}{\end{equation}}
\newcommand{\beqa}{\begin{eqnarray}}
\newcommand{\eeqa}{\end{eqnarray}}

\newcommand{\Bbar}{\,\overline{\!B}{}}
\newcommand{\Dbar}{\,\overline{\!D}{}}
\newcommand{\Kbar}{\,\overline{\!K}{}}
\def\B0bar{\Bbar{}^0}
\def\D0bar{\Dbar{}^0}
\def\K0bar{\Kbar{}^0}

\begin{document}
\begin{flushright}
WSU-HEP-1712
\end{flushright}
\title{\boldmath Radiative lepton flavor violating B, D, and K decays}

\author{Derek Hazard}
\affiliation{Department of Physics and Astronomy\\
        Wayne State University, Detroit, MI 48201, USA}

\author{Alexey A.\ Petrov}
\affiliation{Department of Physics and Astronomy\\
        Wayne State University, Detroit, MI 48201, USA}
\affiliation{Michigan Center for Theoretical Physics\\
        University of Michigan, Ann Arbor, MI 48196, USA}


\begin{abstract}
We argue that radiative lepton flavor violating (RLFV) decays $P \to \gamma \ell_1 \overline{\ell}_2$ of $P =B^0_q$, $\bar{D}^0$, and $K^0$ 
meson states are robust probes of new physics models.  In particular, they could be used to put constraints on the Wilson coefficients of effective 
operators describing lepton flavor-changing neutral current interactions at low energy scales. We set up a generic framework for describing 
these transitions and review new physics constraints from $P \to \ell_1 \bar \ell_2$ decays. There is discussion of how 
RLFV transitions provide access to the operators that cannot be constrained in two-body decays and we in turn motivate further experimental 
searches via these channels.
\end{abstract}

\maketitle

\section{Introduction}\label{Intro}

Currently operating and future B-factories, such as LHCb and Belle-II, will be accumulating significant amounts of 
beauty and charm decay data. These large data sets will be quite useful in studies of extremely small decay rates of 
$B$ and $D$ mesons, which could probe new physics (NP) at unprecedentedly high energy scales. In particular, studies of 
pseudoscalar meson decays $P=B_q^0$, $\bar{D}^0$, and $K^0$ into the final states containing charged leptons of different flavors 
such as $P \to \ell_1 \overline{\ell}_2$ and $P \to \gamma \ell_1 \overline{\ell}_2$ could be performed. Such decays are induced 
by the operators that generate flavor-changing neutral currents (FCNC) in the 
lepton sector, which provide a fruitful approach to probing beyond the standard model (BSM) physics, 
assuming of course that such flavor-violating interactions are allowed in the BSM models. 
There are indeed many well-established new physics models (see, e.g. \cite{Raidal:2008jk,Dreiner:2001kc,Dreiner:2006gu,Sun:2012yq}) 
that meet this opportunity and predict charged lepton flavor violating (CLFV) transition rates that are significantly larger than 
the standard model (SM) rates \cite{Raidal:2008jk}. 

A convenient way to describe CLFV transitions in low energy experiments is by introducing an effective Lagrangian, 
${\cal L}_{\rm eff}$. Such a Lagrangian is a convenient parameterization of all new physics models that include 
lepton flavor violation with the details of the models encoded in the Wilson coefficients (WCs) of ${\cal L}_{\rm eff}$,
which are obtained by matching the effective Lagrangian to a given BSM model at the new physics scale $\Lambda$ 
\cite{Petrov:2016azi}. This Lagrangian is required to be invariant under the unbroken symmetry groups 
$SU(3)_c\times U(1)_{\rm em}$ below the electroweak symmetry breaking scale. At the low scale for which a given 
process occurs the effective operators would exhibit the relevant standard model (SM) degrees of freedom with the effective 
operators written completely using quarks ($q_i = b, c, s, u,$ and $d$) and leptons ($\ell_i = \tau, \mu,$ and $e$). In what 
follows, we assume that top quarks are integrated out of the theory, and we do not consider neutrinos. The effective 
Lagrangian ${\cal L}_{\rm eff}$ that involves CLFV can be written as
\begin{equation}\label{eqn:Leff}
{\cal L}_{\rm eff}= {\cal L}_{\ell q} + {\cal L}_D + \cdot \cdot \cdot,
\end{equation}
where ${\cal L}_D$ is a dipole part, and ${\cal L}_{\ell q}$ is the part that contains four-fermion interactions.
Since here we are interested in the decays of electrically-neutral pseudoscalar $B^0_q$, $\bar{D}^0$, and $K^0$ mesons to
flavor-off-diagonal lepton pairs and other particles, the transitions involve FCNC interactions on both quark and 
lepton sides.  

The dipole part of Eq.~(\ref{eqn:Leff}), which could contribute to the radiative decays $P \to \gamma \ell_1 \overline{\ell}_2$ 
is written as \cite{Celis:2014asa}
\begin{eqnarray}\label{eqn:LD}
{\cal L}_{D} = -\frac{m_2}{\Lambda^2} \left[
\left( 
C_{DR}^{\ell_1\ell_2} \ \overline \ell_1 \sigma^{\mu\nu} P_L \ell_2 + 
C_{DL}^{\ell_1\ell_2} \ \overline \ell_1 \sigma^{\mu\nu} P_R \ell_2 
\right) F_{\mu\nu} + h.c. \right] \text{.}
\end{eqnarray}
The WCs of ${\cal L}_{D}$ have been well constrained in leptonic LFV decays \cite{Raidal:2008jk}.

Note that it is known that the quark FCNC transitions, at least in the decays of down-type quarks,
are dominated by the SM contributions. For instance, the dipole operator describing $q_1 \to q_2 \gamma$
can be written as \cite{Buchalla:1995vs}
\begin{equation}\label{eqn:penguin}
{\cal L}_\text{peng} = \frac{G_F}{\sqrt{2}} \sum_{q} \lambda^P_q C_{7 \gamma} \frac{\sqrt{4 \pi \alpha}}{\pi^2} 
\frac{m_{q_1}}{2} \overline{q}_1 \sigma_{\mu \nu} \left(1 + \gamma_5\right) F^{\mu \nu} q_2 + h.c. 
\end{equation}
Here $\lambda^P_q = V_{q q_2} V_{q q_1}^*$ denotes the appropriate Cabibbo-Kobayashi-Maskawa (CKM) 
matrix elements, $m_{q_1}$ is the heavier quark, and $C_{7 \gamma}$ is the corresponding 
Wilson coefficient \cite{Buchalla:1995vs}. 

The four-fermion dimension-six lepton-quark part of the effective Lagrangian, Eq.~(\ref{eqn:Leff}), takes the 
form \cite{Celis:2014asa}:
\begin{eqnarray}\label{eqn:Llq}
{\cal L}_{\ell q} = -\frac{1}{\Lambda^2} \sum_{q_1, q_2} \Big[
\left( C_{VR}^{q_1 q_2 \ell_1\ell_2} \ \overline\ell_1 \gamma^\mu P_R \ell_2 + 
C_{VL}^{q_1 q_2 \ell_1\ell_2} \ \overline\ell_1 \gamma^\mu P_L \ell_2 \right) \ \overline q_1 \gamma_\mu q_2 &&
\nonumber \\
+ \
\left( C_{AR}^{q_1 q_2 \ell_1\ell_2} \ \overline\ell_1 \gamma^\mu P_R \ell_2 + 
C_{AL}^{q_1 q_2 \ell_1\ell_2} \ \overline\ell_1 \gamma^\mu P_L \ell_2 \right) \ \overline q_1 \gamma_\mu \gamma_5 q_2 &&
\nonumber \\
+ \
m_2 m_{q_{\text{H}}} G_F \left( C_{SR}^{q_1 q_2 \ell_1\ell_2} \ \overline\ell_1 P_L \ell_2 + 
C_{SL}^{q\ell_1\ell_2} \ \overline\ell_1 P_R \ell_2 \right) \ \overline q_1 q_2 &&
\\
+ \
m_2 m_{q_{\text{H}}} G_F \left( C_{PR}^{q_1 q_2 \ell_1\ell_2} \ \overline\ell_1 P_L \ell_2 + 
C_{PL}^{q_1 q_2 \ell_1\ell_2} \ \overline\ell_1 P_R \ell_2 \right) \ \overline q_1 \gamma_5 q_2 
\nonumber \\
+ \
m_2 m_{q_{\text{H}}}G_F \left( C_{TR}^{q_1 q_2 \ell_1\ell_2} \ \overline\ell_1 \sigma^{\mu\nu} P_L \ell_2 + 
C_{TL}^{q_1 q_2 \ell_1\ell_2} \ \overline\ell_1 \sigma^{\mu\nu} P_R \ell_2 \right) \ \overline q_1 \sigma_{\mu\nu} q_2 
 &+& h.c. ~ \Big] .
\nonumber
\end{eqnarray}
Here $m_{q_{\text{H}}}$ is the mass of the heavier quark ($m_{q_{\text{H}}} = \text{max}[m_{q_1},  m_{q_2}]$) and 
$P_{\rm R,L}=(1\pm \gamma_5)/2$ is the right (left) chiral projection operator. In general the Wilson coefficients would
be different for different lepton flavors $\ell_i$ and quark flavors $q_i$. Note that, contrary to some previous studies,
we include tensor operator in Eq.~(\ref{eqn:Llq}) (see \cite{Hazard:2016fnc} for motivation). CP-conservation is assumed 
so all the Wilson coefficients in Eq.~(\ref{eqn:Llq}) should be viewed as real numbers.

In this paper we discuss the possibility of the Wilson coefficients of the effective Lagrangian in
Eq.~(\ref{eqn:Leff}) for different $\ell_i$ and $q_i$ be determined from experimental data on 
leptonic and radiative leptonic CLFV decays of $B^0_q$, $\bar{D}^0$, and $K^0$ states.
We review two-body decays $P \to \ell_1 \overline{\ell}_2$ in Sect.~\ref{Spin0LLgam}. We will note that 
restricted kinematics of the two-body transitions would allow us to select operators with particular 
quantum numbers significantly reducing the reliance on the single operator dominance assumption \cite{Hazard:2016fnc}.
The main part of the paper, Sect.~\ref{Spin0LLgam},  will be devoted to discussion of radiative lepton-flavor violating (RLFV) decays
$P \to \gamma \ell_1 \overline{\ell}_2$. We will summarize our results in Sect.~\ref{Results} and conclude in Sect.~\ref{Conclusions}.

Note that here we only consider short distance effects in Kaon decays.  In the SM long distance effects on decays such as
$K_{L(S)}^0 \to \gamma \ell \bar \ell$ dominate the dynamics \cite{Cirigliano:2011ny}.  In light of this, our kaon results may
be modified by long distance effects.

In what follows, we will use the convention that the subscript of ``$1$'' will denote the lighter 
lepton and the subscript ``$2$'' will denote the heavier lepton.  Unless otherwise specified when studying 
the branching ratios we assume for a meson, $P$, that ${\cal B}\left(P \to \left(\gamma\right) \ell_1 \ell_2 \right) 
= {\cal B}\left(P \to \left(\gamma\right) \overline \ell_1\ell_2 \right) +{\cal B}\left(P \to \left(\gamma\right) \ell_1 
\overline \ell_2 \right)$.  Finally, it is important to note that some of the two-body and all of 
the three-body transitions have yet to be experimentally studied. Numerical constraints on some Wilson coefficients of the effective 
Lagrangian, ${\cal L}_{eff}$, from these unstudied decays are not available.

\section{Two-body decays $P \to \ell_1 \overline \ell_2$}\label{Spin0LLgam}

Many studies have focused on rare leptonic decays of $B_q^0$ mesons, $B_q \to \ell \bar \ell$, as both precision tests of the SM and 
as an opportunity to search for new physics (e.g.  \cite{Altmannshofer:2017wqy,Beneke:2017vpq,Blake:2016olu,Bobeth:2013uxa,Bobeth:2013tba}).
The abundance of produced $B_q^0$ and $\bar{D}^0$ states at the LHCb, Belle II, and BESIII experiments also allows for 
studies of lepton-flavor violating decays at these experiments \cite{Amhis:2016xyh,PDG}. Such decays were discussed at length 
previously, mainly in the context of particular models. Here we shall review these transitions emphasizing the 
possibility to constrain Wilson coefficients of the axial and pseudoscalar operators of the effective Lagrangian in Eq.~(\ref{eqn:Leff}). 
These decays would provide information about $C_{PL}^{q_1 q_2 \ell_1\ell_2}(C_{PR}^{q_1 q_2 \ell_1\ell_2})$ and/or 
$C_{AL}^{q_1 q_2 \ell_1\ell_2}(C_{AR}^{q_1 q_2 \ell_1\ell_2})$ 
in Eq.~(\ref{eqn:Llq}). 
\begin{table*}
\caption{\label{tab:Pdecaylimits} Available experimental limits on ${\cal B}(P \to \ell_1 \overline{\ell}_2)$ \cite{PDG, Aaij:2017cza, Aubert:2008cu, Aaij:2015qmj, Ambrose:1998us}.
Center dots signify that no experimental data are available; ``FPS" means that the transition is forbidden by phase space.}
\begin{ruledtabular}
\begin{tabular}{cccc}
$\ell_1 \ell_2$& $\mu \tau$ & $e \tau$ & $e \mu$  \\ 
\hline
${\cal B}(B^0_d \to \ell_1 \ell_2)$ & $2.2 \times 10^{-5}$ & $2.8 \times 10^{-5}$ & $1.0 \times 10^{-9}$ \\ 
${\cal B}(B^0_s \to \ell_1 \ell_2)$ & $\cdot \cdot \cdot$ & $\cdot \cdot \cdot$ & $5.4 \times 10^{-9}$ \\
${\cal B}(\bar{D}^0 \to \ell_1 \ell_2)$ & FPS & $\cdot \cdot \cdot$ & $1.3 \times 10^{-8}$ \\
${\cal B}(K^0_L \to \ell_1 \ell_2)$ & FPS & FPS & $4.7 \times 10^{-12}$ \\
\end{tabular}
\end{ruledtabular}
\end{table*}
One can write the most general expression for the $P \to \ell_1 \overline \ell_2$ decay amplitude as \cite{Hazard:2016fnc}
\begin{eqnarray}\label{Spin0Amp}
{\cal A}(P\to \ell_1 \overline \ell_2) = \overline{u}(p_1, s_1) \left[
E_P^{q_1 q_2 \ell_1\ell_2}  + i F_P^{q_1 q_2 \ell_1\ell_2} \gamma_5 
\right] v(p_2,s_2) \,
\end{eqnarray}
with $E_P^{q_1 q_2 \ell_1\ell_2}$ and $F_P^{q_1 q_2 \ell_1\ell_2}$ being dimensionless constants which 
depend on the Wilson coefficients of operators in Eq.~(\ref{eqn:Leff}) and various decay constants.

The amplitude of Eq.~(\ref{Spin0Amp}) leads to the branching ratio for flavor off-diagonal leptonic 
decays of pseudoscalar mesons:
\begin{eqnarray}\label{BRSpin0}
{\cal B}(P \to \ell_1 \overline \ell_2) = \frac{m_P}{8\pi \Gamma_P} \left(1-y^2\right)^2
\left[\left|E_P^{q_1 q_2\ell_1\ell_2}\right|^2 + \left|F_P^{q_1 q_2\ell_1\ell_2}\right|^2\right].
\end{eqnarray}
Here $\Gamma_P$ is the total width of the pseudoscalar state.  We have once again 
neglected the mass of the lighter lepton and set $y = m_2/m_P$. 
Calculating $E_P^{q_1 q_2 \ell_1\ell_2}$ and $F_P^{q_1 q_2 \ell_1\ell_2}$  for
 $P=B^0_d$ ($q_1 q_2 = d b$), $B^0_s$ ($q_1 q_2 = s b$), $\bar{D}^0$ ($q_1 q_2=cu$), and , $K^0_L$ ($q_1 q_2= d s$), the coefficients are
\begin{align}\label{PCoef1}
\begin{split}
E_P^{q_1 q_2\ell_1\ell_2} &= \kappa_P \frac{m_P f_{P} y}{2 \Lambda^2} \left[ \left(C_{AL}^{q_1 q_2 \ell_1\ell_2}+C_{AR}^{q_1 q_2 \ell_1\ell_2}\right) + 
m_{P}^{2} G_{F} \left( C_{PL}^{q_1 q_2 \ell_1\ell_2} + C_{PR}^{q_1 q_2 \ell_1\ell_2} \right) \right], \\
F_P^{q_1 q_2\ell_1\ell_2} &= i \kappa_P \frac{m_P  f_{P} y}{2 \Lambda^2}\left[\left(C_{AL}^{q_1 q_2 \ell_1\ell_2}-C_{AR}^{q_1 q_2 \ell_1\ell_2}\right) + 
m_P^2 G_F \left(C_{PL}^{q_1 q_2 \ell_1\ell_2}-C_{PR}^{q_1 q_2 \ell_1\ell_2}\right) \right].
\end{split}
\end{align}
The hadronic matrix element in Eq.~(\ref{PCoef1}) is defined as \cite{Petrov:2013vka}
\begin{eqnarray}\label{DeConP}
&& \langle 0| \overline q_1 \gamma^\mu \gamma_5 q_2 | P(p) \rangle = -i f_P p^\mu\,.
\end{eqnarray}
Here $p$ is the momentum of the meson. The constant $\kappa_P$ is $1$ for $B_q^0$, $\bar{D}^0$, and $K^0$; and $1/\sqrt{2}$ for $K^0_{L(S)}$.
The experimental limits and numerical values of the pseudo-scalar decay constants used in the calculations can be found in Tables 
\ref{tab:Pdecaylimits} and \ref{tab:Pdecayconstants}. The resulting constraints on the Wilson coefficients are found in Table \ref{tab:ps_constr}.

\begin{table*}
\caption{\label{tab:Pdecayconstants} Pseudoscalar meson decay constants \cite{Dowdall:2013tga,Carrasco:2014poa}, 
total decay widths, and meson masses \cite{PDG} used in the calculation of branching ratios ${\cal B}(P \to \ell_1 \overline \ell_2)$.}
\begin{ruledtabular}
\begin{tabular}{|c|cccc|}
~State & $B^0_d$ & $B^0_s$ & $\bar{D}^0$ & $K^0_L$ \\
\hline
~$f_P$, MeV~ &  $186 \pm 4$   &  $224 \pm 4$ &  $207.4 \pm 3.8$ &  $155.0 \pm 1.9$ \\
~$\Gamma_{P}$, $10^{-14}$ MeV & $4330 \pm 11$ & $4374 \pm 15$ & $16050 \pm 60$ & $1.287 \pm 0.005$ \\
~$m_P$, GeV~ & $5.28$ & $5.37$ & $1.86$ & $0.498$ \\
\end{tabular}
\end{ruledtabular}
\end{table*} 
\begin{table*}
\caption{\label{tab:ps_constr}Constraints on the Wilson coefficients from pseudoscalar meson decays.  Note the $K^0_L$ 
results only include short distance effects. Center dots signify that no experimental data are available to produce a 
constraint; ``FPS" means that the transition is forbidden by phase space. 
Particle masses and other input parameters are from \cite{PDG, Aaij:2017cza, Aubert:2008cu, Aaij:2015qmj, Ambrose:1998us}.}
\begin{ruledtabular}
\begin{tabular}{cccccc}
 & Leptons &\multicolumn{4}{c}{Initial state}\\
 Wilson coefficient & $\ell_1 \ell_2$ & $B^0_d \left(d \bar b\right)$ & $B^0_s\left(s \bar b\right)$ & $\bar{D}^0 \left(u \bar c \right)$ & $K^0_L \left(\left(d \bar s - s \bar d \right)/\sqrt{2}\right)$ \\ \hline
$\left| {C_{AL}^{q_1 q_2 \ell_1\ell_2}}/{\Lambda^2} \right|$ & $\mu \tau$ & $2.3 \times 10^{-8}$ &  $\cdot \cdot \cdot$  & FPS & FPS \\
$~$ & $e \tau$ & $2.6 \times 10^{-8}$ &  $\cdot \cdot \cdot$  &  $\cdot \cdot \cdot$  & FPS  \\
$~$ & $e \mu$ & $2.3 \times 10^{-9}$ & $4.4 \times 10^{-9}$ & $2.4 \times 10^{-8}$ & $5.0 \times 10^{-12}$ \\
\hline
$\left| {C_{AR}^{q_1 q_2 \ell_1\ell_2}}/{\Lambda^2} \right|$ & $\mu \tau$ & $2.3 \times 10^{-8}$ &  $\cdot \cdot \cdot$  & FPS & FPS \\
$~$ & $e \tau$ & $2.6 \times 10^{-8}$ &  $\cdot \cdot \cdot$  &  $\cdot \cdot \cdot$  & FPS \\
$~$ & $e \mu$ & $2.3 \times 10^{-9}$ & $4.4 \times 10^{-9}$ & $2.4 \times 10^{-8}$ & $5.0 \times 10^{-12}$ \\
\hline
$\left| {C_{PL}^{q_1 q_2 \ell_1\ell_2}}/{\Lambda^2} \right|$ & $\mu \tau$ & $7.1 \times 10^{-5}$ &  $\cdot \cdot \cdot$  & FPS & FPS \\
$~$ & $e \tau$ & $8.0 \times 10^{-5}$ &  $\cdot \cdot \cdot$  &  $\cdot \cdot \cdot$  & FPS \\
$~$ & $e \mu$ & $7.1 \times 10^{-6}$ & $1.3 \times 10^{-5}$ & $5.9 \times 10^{-4}$ & $1.7 \times 10^{-6}$ \\
\hline
$\left| {C_{PR}^{q_1 q_2 \ell_1\ell_2}}/{\Lambda^2} \right|$ & $\mu \tau$ & $7.1 \times 10^{-5}$ &  $\cdot \cdot \cdot$  & FPS & FPS \\
$~$ & $e \tau$ & $8.0 \times 10^{-5}$ &  $\cdot \cdot \cdot$  &  $\cdot \cdot \cdot$  & FPS \\
$~$ & $e \mu$ & $7.1 \times 10^{-6}$ & $1.3 \times 10^{-5}$ & $5.9 \times 10^{-4}$ & $1.7 \times 10^{-6}$ \\
\end{tabular}
\end{ruledtabular}
\end{table*}

\section{Three-body radiative decays $P \to \overline \ell_1 \ell_2\gamma$}\label{Spin0LLgam}

Similarly to the $B^0_s \to \mu^+\mu^- \gamma$ transition
\cite{Aditya:2012im,Kruger:2002gf,Melikhov:2017pwu,Kozachuk:2016ypz,Melikhov:2004mk}, addition of a photon 
to the $\ell_1 \overline{\ell}_2$ final state allows one to probe operators of the effective Lagrangian that 
do not contribute to $P\to \ell_1 \overline{\ell}_2$ transition. This was pointed out for the LFV 
decays in \cite{Hazard:2016fnc}, and, more importantly in \cite{Guadagnoli:2016erb} (for a calculation of 
$B^0_s \to  \ell_1 \overline{\ell}_2 \gamma$ in the model of \cite{Glashow:2014iga}). In addition, $P \to \ell_1 \bar \ell_2$ decays suffer
from chiral suppression (see Eq. (\ref{PCoef1})), which three-body radiative decays do not neccessarily exhibit.  Thus,
it is possible that RLFV decays might have larger branching ratios than two-body LFV transitions 
(see \cite{Aditya:2012im,Kruger:2002gf,Melikhov:2017pwu,Kozachuk:2016ypz,Melikhov:2004mk} for similar effects in lepton
flavor conserving decays).  Here we evaluate radiative lepton-flavor 
violating decays of the pseudoscalar mesons with the model-independent effective Lagrangian of Eq.~(\ref{eqn:Leff}).

It might be theoretically easier to deal with a three-body final state that contains no strongly-interacting composite particles.
Still, the calculation of the $P\to \ell_1 \overline{\ell}_2 \gamma$ decay is more complicated than $P\to \ell_1 \overline{\ell}_2$, where all 
nonperturbative effects are summarized in one decay constant $f_P$. Further, because of the electromagnetic gauge 
invariance, it is important to have a good understanding of what kind of constraints the kinematic structure of the decay 
amplitude imposes on the dynamics of these transitions. Let us now derive the most general amplitude for 
$P\to \ell_1 \overline{\ell}_2 \gamma$.

\subsection{General amplitude and differential decay rate for $P \to \overline \ell_1 \ell_2\gamma$}\label{Amplitude}

The most general expression for the $P(p) \to \gamma (k) \ell_1(p_1) \overline \ell_2 (p_2)$ decay amplitude 
can be obtained using the Bardeen-Tung formalism \cite{Bardeen:1969aw}. The decay amplitude can be written 
as
\begin{eqnarray}\label{PGLLAmp1}
A(P(p) \to \gamma (k) \ell_1(p_1) \overline \ell_2 (p_2)) = \overline{u}(p_1, s_1) \ 
M^\mu (p,k,q) \ v(p_2,s_2) \ \varepsilon^*_\mu(k),
\end{eqnarray}
where $\overline{u}(p_1, s_1)$ and $v(p_2,s_2)$ are spinors for $\ell_1$ and $\bar \ell_2$, $q=\frac{1}{2}\left(p_1-p_2\right)$, and 
 $\varepsilon^*_\mu(k)$ is the polarization vector of the photon. The function $M^\mu (p,k,q)$, which we seek to parameterize, 
transforms as a tensor under Lorentz transformations. This function should only contain dynamical singularities, so particular 
care should be taking by writing it in such a way that it does not contain kinematical ones. The most general expression for the 
$M^\mu (p,k,q)$ from Eq.~(\ref{PGLLAmp1}) can be written by  expanding it into simpler Lorentz structures $\ell_i^\mu(p,q,k)$ multiplied by 
the invariant functions $M_i^{P\ell_1\ell_2}$, which only depend on Lorentz invariants, 
\begin{equation}\label{AtoM}
M^\mu (p,k,q) = \sum_i \ell_i^\mu(p,q,k) M_i^{P\ell_1\ell_2} (p^2, ...) \ .
\end{equation}
The most general parameterization of Eq.~(\ref{AtoM}) contains twelve form-factors,
\begin{eqnarray}\label{M1}
M^\mu (p,k,q) &=&  \gamma^\mu \left(M_1^{P\ell_1\ell_2} + \slashed{k} M_2^{P\ell_1\ell_2}\right) 
+ i \gamma_5 \gamma^\mu \left(M_3^{P\ell_1\ell_2} + \slashed{k} M_4^{P\ell_1\ell_2}\right) 
\nonumber \\
&+&  q^\mu \left(M_5^{P\ell_1\ell_2} + \slashed{k} M_6^{P\ell_1\ell_2}\right) 
+ i \gamma_5 q^\mu \left(M_7^{P\ell_1\ell_2} + \slashed{k} M_8^{P\ell_1\ell_2}\right) 
\\
&+&  p^\mu \left(M_9^{q\ell_1\ell_2} + \slashed{k} M_{10}^{q\ell_1\ell_2}\right) 
+ i \gamma_5 p^\mu \left(M_{11}^{P\ell_1\ell_2} + \slashed{k} M_{12}^{P\ell_1\ell_2}\right) \ .
\nonumber
\end{eqnarray}
In writing of Eq.~(\ref{M1}) we used the equation of motion for the lepton spinors, and rewrote terms containing $\sigma^{\mu\nu}$ 
in terms of components, e.g. $i \sigma^{\mu\nu} q_\nu = q^\mu-\gamma^\mu \slashed{q}$. Note that terms proportional to 
$\slashed{q}$ can be expressed as terms proportional to $\slashed{k}$ using momentum conservation and equations of 
motion. Next, terms proportional to the 
$\epsilon^{\mu\nu\alpha\beta}$ tensor, such as $\epsilon^{\mu\nu\alpha\beta}\gamma_\nu p_\alpha k_\beta$, can be 
written in terms of the existing form factors of Eq.~(\ref{M1}) using the relation
\begin{equation}
i \epsilon^{\mu\nu\alpha\beta}\gamma_\beta = 
\gamma^\mu\gamma^\nu\gamma^\alpha\gamma_5 - g^{\mu\nu} \gamma^\alpha \gamma_5 -
g^{\nu\alpha} \gamma^\mu \gamma_5 + g^{\mu\alpha} \gamma^\nu \gamma_5
\end{equation}
and the equations of motion. Finally, all possible terms in Eq.~(\ref{M1}) proportional to $k^\mu$ trivially vanish by gauge invariance. 

The set of Eq.~(\ref{M1}) is still not minimal, as the condition of gauge invariance 
$ k_\mu M^\mu (p,k,q)=0$ implies that some of the $M_i^{P\ell_1\ell_2}$ in Eq.~(\ref{M1}) are not independent. An elegant way of 
finding the minimal set of gauge-invariant Lorentz structures has been given in \cite{Bardeen:1969aw}, which we shall apply
to our analysis. To get the minimal set, it is most convenient to apply a projection operator  
\begin{equation}\label{Pmunu}
P^{\mu\nu} = g^{\mu\nu} - \frac{p^\mu k^\nu}{(p\cdot k)} 
\end{equation}
to $M^\mu (p,k,q)$. Since $P^{\mu\nu} M_\nu = M^\mu$ and $k_\mu P^{\mu\nu}=0$, $P^{\mu\nu}$ 
does indeed project out gauge-invariant structures in $M^\mu (p,k,q)$. Applying $P^{\mu\nu}$ to Eq.~(\ref{M1}) 
we learn that terms proportional to $p^\mu$ do not give contributions to the minimal set and should be dropped, 
leaving the number of independent amplitudes at eight. Applying the condition $k_\mu \ell_i^\mu=0$ and eliminating 
kinematical singularities we write the Lorentz structures  $L^\mu_i$ for the set of amplitudes as

%
\begin{equation}\label{AtoMfinal}
M^\mu (p,k,q) = \sum_i L_i^\mu(p,q,k) A_i^{P\ell_1\ell_2} (p^2, ...) \text{,}
\end{equation}
which are defined in a manner that removes all kinematical singularities. The $A_i^{P\ell_1\ell_2} (p^2, ...)$ are new scalar 
form factors, while $L^\mu_i$ are
\begin{eqnarray}\label{Ls}
L_1^\mu &=& \gamma^\mu \slashed{k}, ~~ L_2^\mu = i \gamma_5 \gamma^\mu \slashed{k}, 
\nonumber \\
L_3^\mu &=& \left(p\cdot k\right) q^\mu -  \left(k\cdot q\right) p^\mu,
\nonumber \\
 L_4^\mu &=& i \gamma_5\left[\left(p\cdot k\right) q^\mu -  \left(k\cdot q\right) p^\mu \right],
\nonumber \\
L_5^\mu &=& \left(p\cdot k\right) \gamma^\mu -  p^\mu \slashed{k},
\\
L_6^\mu &=& i \gamma_5\left[\left(p\cdot k\right) \gamma^\mu -  p^\mu \slashed{k} \right],
\nonumber \\
L_7^\mu &=& q^\mu \slashed{k} - \left(k\cdot q\right) \gamma^\mu,
\nonumber \\
L_8^\mu &=& i \gamma_5\left[q^\mu \slashed{k} - \left(k\cdot q\right) \gamma^\mu \right].
\nonumber
\end{eqnarray}
This implies that the decay amplitude can be written as 
\begin{eqnarray}\label{PGLLAmp2}
A(P(p) \to \gamma (k) \ell_1(p_1) \overline \ell_2 (p_2)) = \sum_i  A_i^{P\ell_1\ell_2} (p^2, ...)  \ \overline{u}(p_1, s_1) \ 
L_i^\mu(p,q,k) \ v(p_2,s_2) \ \varepsilon^*_\mu(k) .
\end{eqnarray}
Using this general amplitude for a three-body pseudoscalar decay, $P \to \gamma \ell_1 \overline{\ell}_2$, we calculate a general differential decay rate, 
which depends on the same scalar functions $A_i^{P\ell_1\ell_2} (p^2, ...)$, 
\begin{eqnarray}
\begin{split} \label{eqn:DiffDecayRate}
\frac{d\Gamma}{dm_{12}^2 dm_{23}^2} = \frac{1}{(2\pi)^3}&\frac{1}{384 m_P^3}\Big[-16\left(A_1^2+A_2^2\right)\left(m_{13}^2\left(m_P^2y^2-m_{23}^2\right)+m_{\gamma}^2 m_P^2 \left(1-y^2\right)\right) \\
+\ 2 &\left(A_3^2+A_4^2\right) \left(m_P^2y^2-m_{12}^2\right) \Big\lbrace m_{13}^2\left(m_P^4y^2-m_{12}^2m_{23}^2\right) \\
&\quad\quad\quad\quad\quad\quad + m_{\gamma}^2 \left(m_{13}^2 m_{23}^2 - \tfrac{1}{4} \left(m_P^2-m_{12}^2+m_{\gamma}^2 \right)^2\right) \Big\rbrace \\ 
+\ 4&\left(A_5^2+A_6^2\right)\Big\lbrace2m_P^6y^4+m_{12}^2\left(\left(m_P^2y^2-m_{13}^2\right)^2+m_{23}^4\right) \\
&\quad\quad\quad\quad\quad\quad-m_P^2y^2\left(m_P^2+m_{12}^2\right)\left(m_P^2y^2+m_{23}^2-m_{13}^2\right)\Big\rbrace \\
-\ \ &\left(A_7^2+A_8^2\right)\Big\lbrace\left(2m_P^2y^2-m_{12}^2\right)\left(\left(m_P^2y^2-m_{23}^2\right)^2+m_{13}^4\right) \\
&\quad\quad\quad\quad\quad\quad+m_P^2y^2\left(m_P^2-m_{12}^2\right)\left(m_P^2y^2-m_{23}^2+m_{13}^2\right)\Big\rbrace  \\
-\ 8&Re\left[A_1A_3^*+A_2A_4^*\right]  \Big\lbrace m_{13}^2 \left(m_P^4y^2-m_{12}^2m_{23}^2\right) \\
&\quad\quad\quad\quad\quad\quad - \tfrac{1}{2} m_{\gamma}^2 \left(m_P^2+m_{\gamma}^2-m_{12}^2\right) \left(m_P^2y^2-m_{12}^2\right) \Big\rbrace \\
-16&Re\left[A_1A_5^*+A_2A_6^*\right]m_Pym_{13}^2\left(m_P^2-m_{12}^2\right) \\
+\ 8&Re\left[A_1A_7^*+A_2A_8^*\right]m_Pym_{13}^2\left(m_P^2y^2-m_{23}^2+m_{13}^2\right) \\
+\ 8&Re\left[A_3A_5^*+A_4A_6^*\right]m_Pym_{13}^2\left(m_P^4y^2-m_{12}^2m_{23}^2\right) \\
+\ 4&Re\left[A_3A_7^*+A_4A_8^*\right]m_Pym_{13}^2\left(m_P^4y^2-m_{12}^2m_{23}^2\right) \\
+\ 4&Re\left[A_5A_7^*+A_6A_8^*\right]\left(m_P^2-m_{12}^2\right)\left(m_P^2y^2-m_{12}^2\right)\left(m_P^2y^2-m_{23}^2+m_{13}^2\right)\Big] \text{.}
\end{split}
\end{eqnarray}
Here the Mandelstam variables have the usual definitions: $m_{12}^2 = (p_1+p_2)^2$, $m_{13}^2 = (p_1+k)^2$, $m_{23}^2 = (p_2+k)^2$, 
where $p_{1,2}$ is the $\ell_{1,2}$ lepton momentum, $k$ is the $\gamma$ photon momentum, and they are related to the pseudoscalar 
momentum, $p$, by $p=p_1+p_2+k$.  The mass $m_P$ is the pseudo-scalar mass, $m_2$ is the heavier lepton mass, and $y=m_2/m_P$.  
The superscript of $P \ell_1 \ell_2$ on the scalar functions $A_i^{P \ell_1 \ell_2}(p^2, ...)$ is dropped for brevity in Eq. (\ref{eqn:DiffDecayRate}).
We introduce a photon mass, $m_{\gamma}$, to regulate the infrared divergences that will appear via bremsstrahlung diagrams.  We use a value of 
$m_{\gamma} = 60$ MeV as our cut-off, which is near the final state invariant mass resolution of experiments \cite{Guadagnoli:2016erb}.

\subsection{Scalar functions $A_i^{P\ell_1\ell_2}$ for $B^0_q$, $\bar{D}^0$, and $K^0$ mesons} \label{ScalarFunctions}

\begin{figure}
\subfigure[]{\includegraphics[scale=0.5]{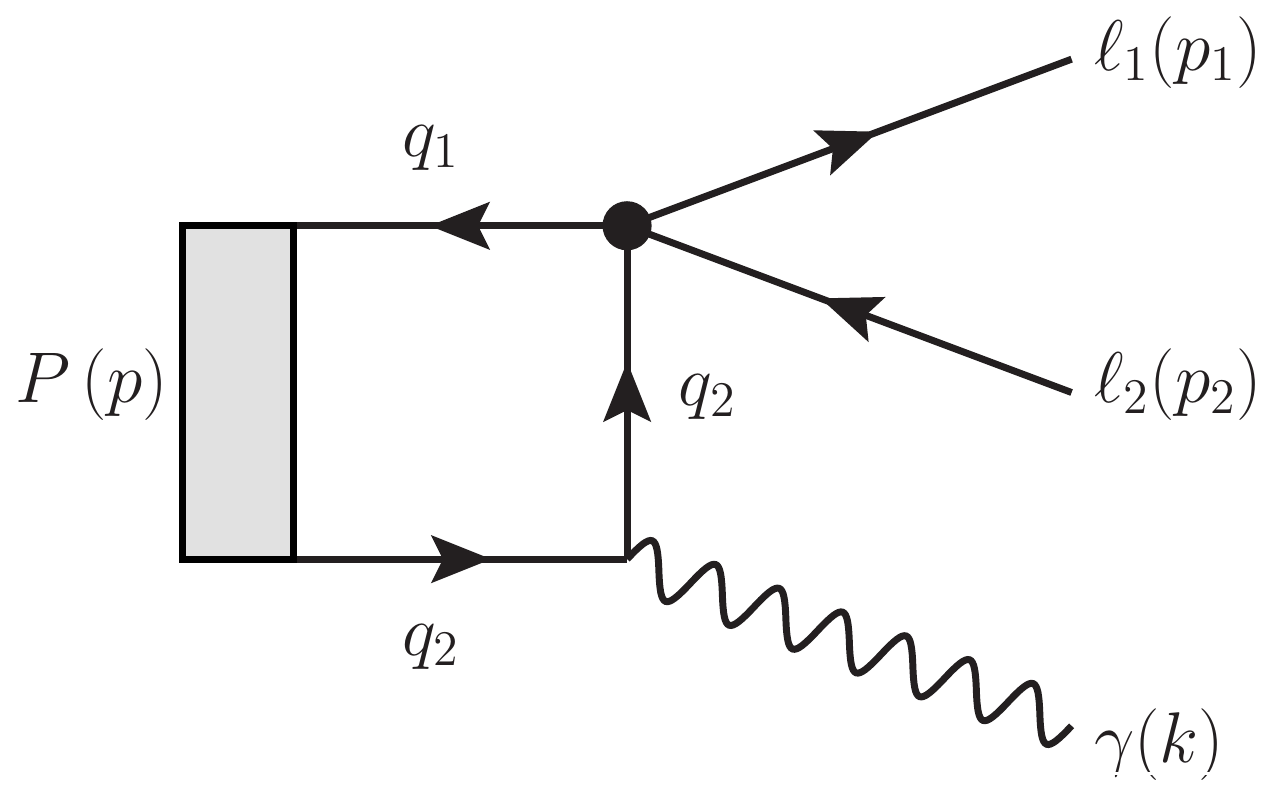} \label{diagram a}}
\hspace{0.5in}
\subfigure[]{\includegraphics[scale=0.5]{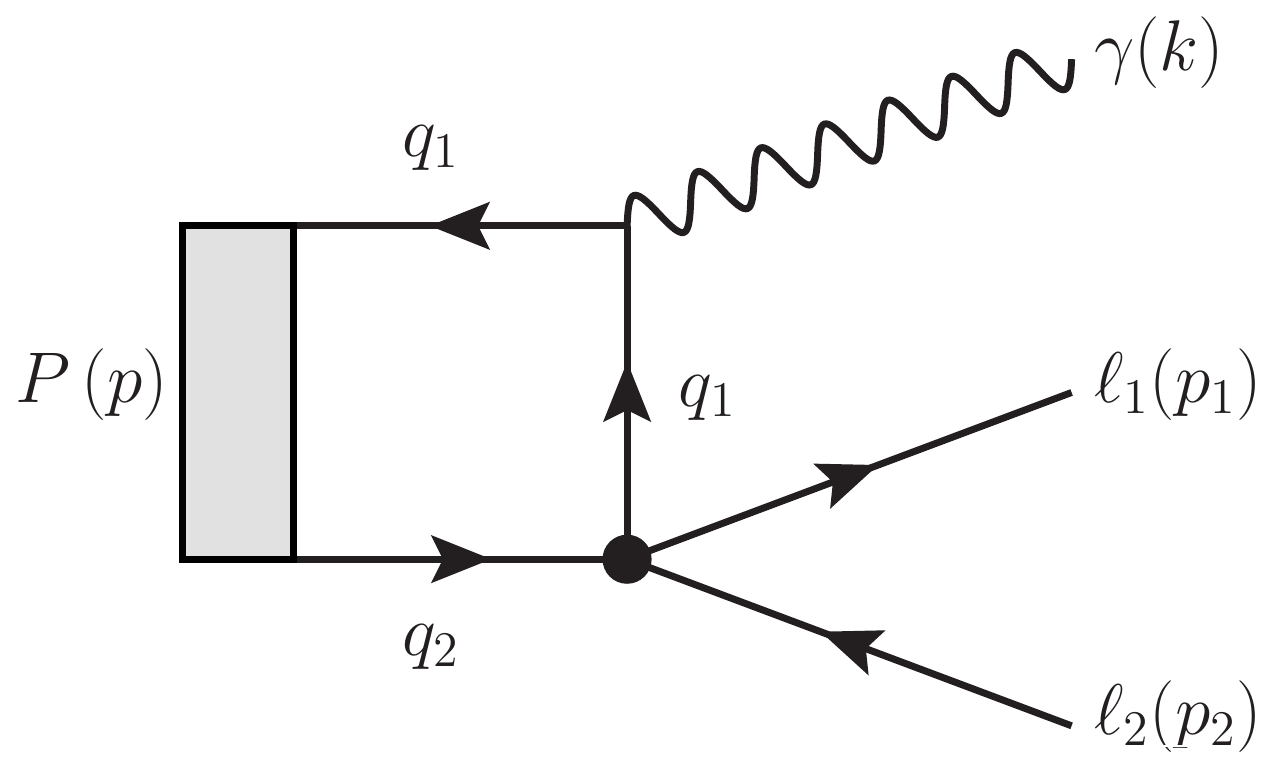} \label{diagram b}}
\caption{ Four-fermion interaction diagrams for ${\cal A} (P \to \gamma \ell_1 \overline \ell_2)$ for operators of type ${\cal O} \sim (\ell_1 \overline{\ell}_2) (\overline{q}_1 q_2)$ where $q_1 \neq q_2$ with photon $\gamma (k)$ attached to the valence quark.  The black circles represent the four-fermion LFV vertex defined in ${\cal L}_{eff}$ of Eq. (\ref{eqn:Llq}).}
\label{3bodydecaydiagramsAB}
\end{figure}

\begin{figure}
\subfigure[]{\includegraphics[scale=0.5]{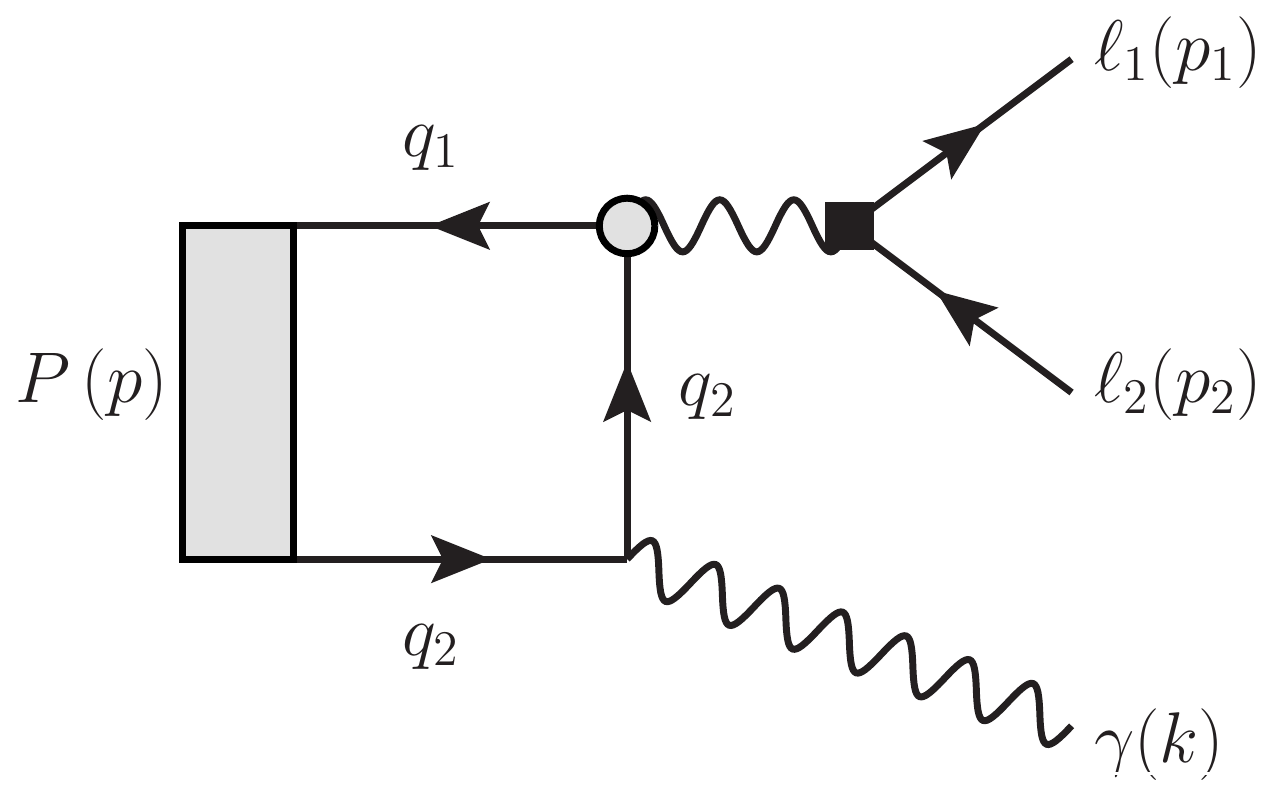} \label{diagram e}}
\hspace{0.5in}
\subfigure[]{\includegraphics[scale=0.5]{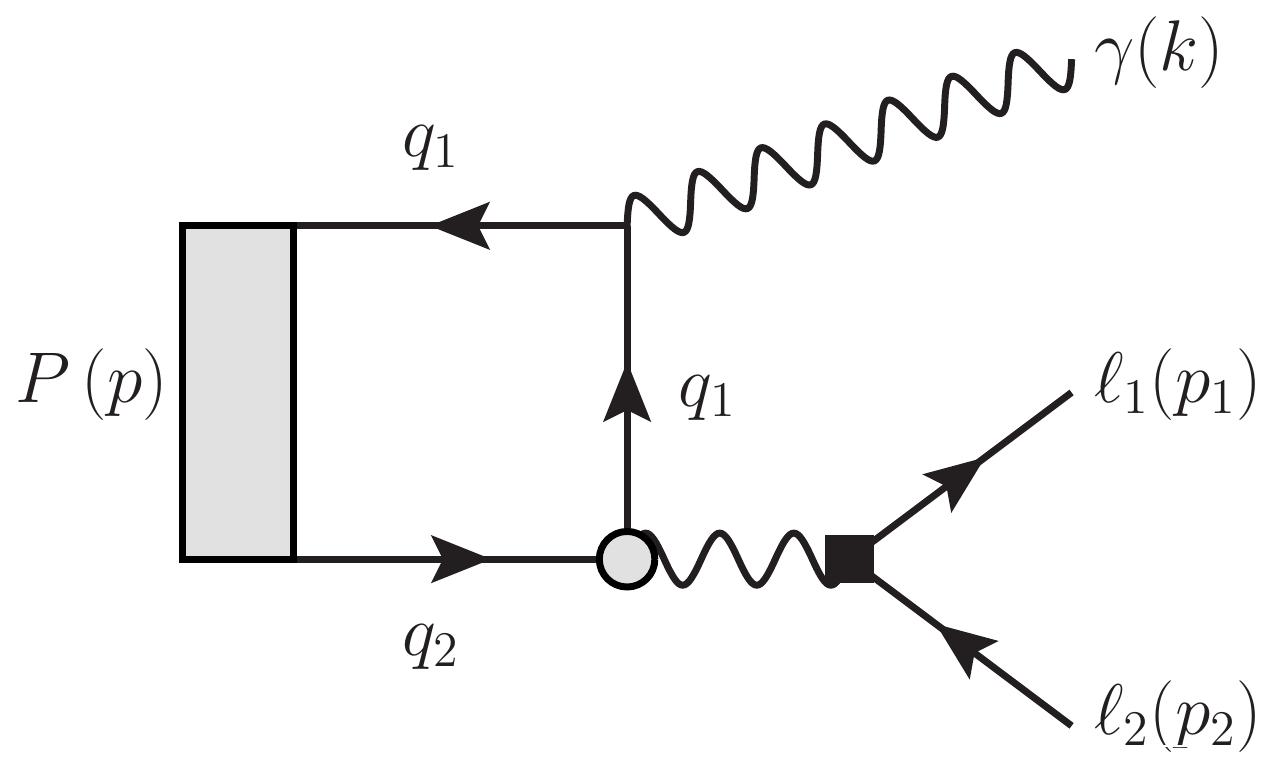} \label{diagram f}} \\
\subfigure[]{\includegraphics[scale=0.5]{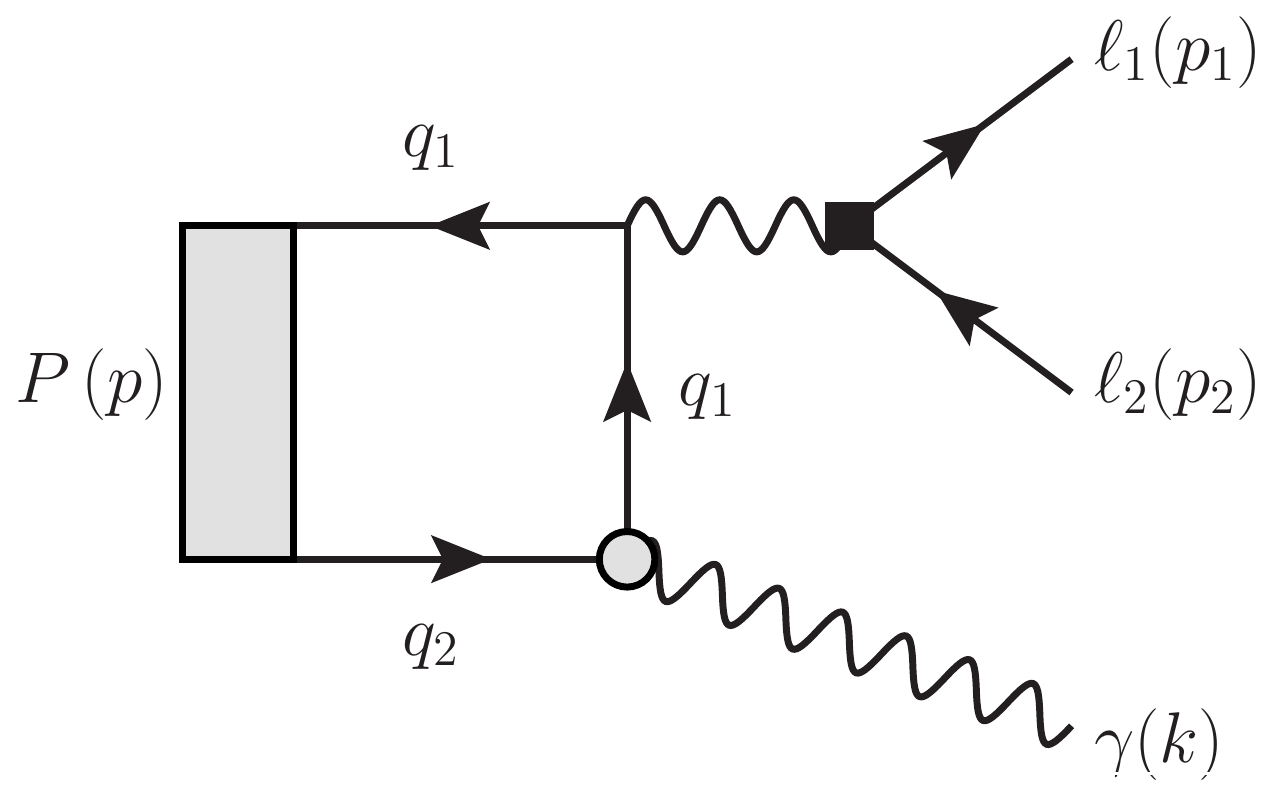} \label{diagram g}}
\hspace{0.5in}
\subfigure[]{\includegraphics[scale=0.5]{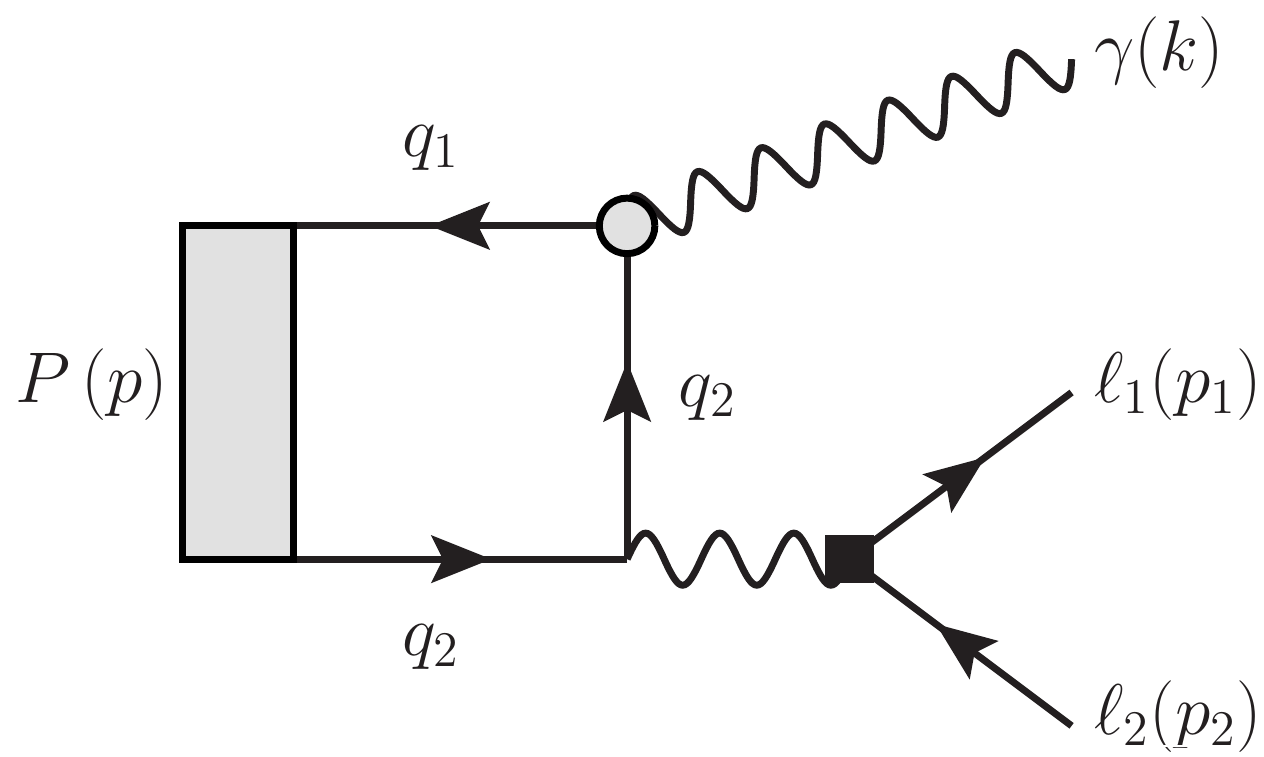} \label{diagram h}}
\caption{ Dipole operator diagrams for ${\cal A} (P \to \gamma \ell_1 \overline \ell_2)$.  The grey circles with the black border represent the SM dipole penguin vertex (Eq. (\ref{eqn:penguin})) and the black boxes represent the dipole LFV vertex (Eq. (\ref{eqn:LD})). Note that the contributions of these diagrams are severely constrained by already available data on $\ell_1 \to \ell_2 \gamma$ decays.}
\label{3bodydecaydiagramsEFGH}
\end{figure}

\begin{figure}
\subfigure[]{\includegraphics[scale=0.5]{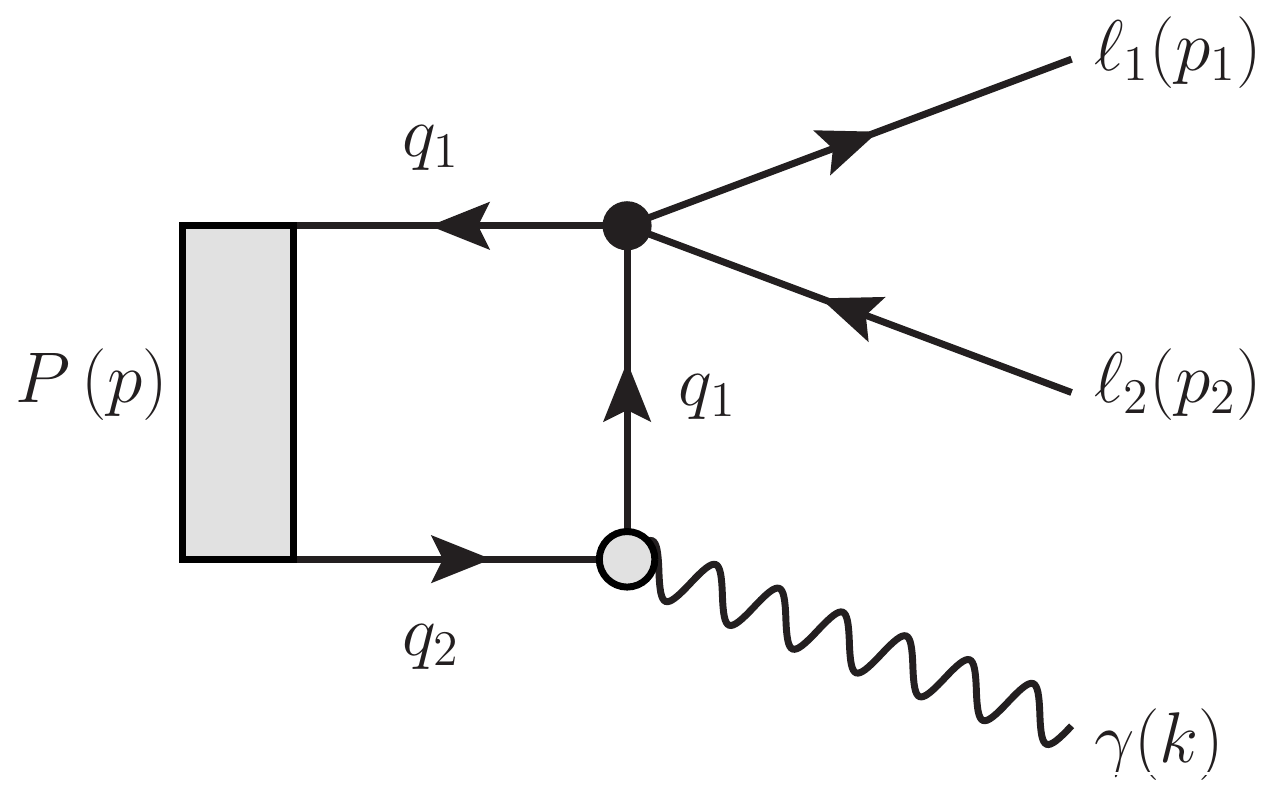} \label{diagram c}}
\hspace{0.5in}
\subfigure[]{\includegraphics[scale=0.5]{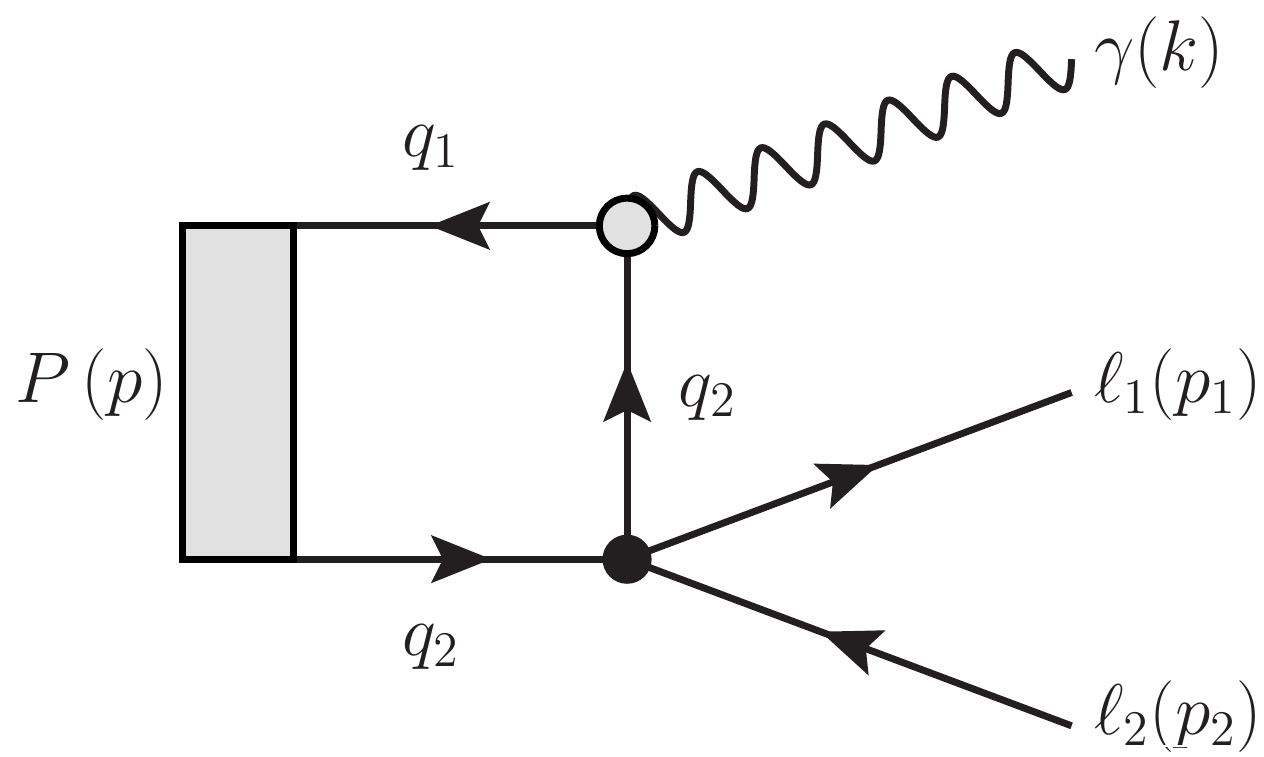} \label{diagram d}}
\caption{Four-fermion interaction diagrams for ${\cal A} (P \to \gamma \ell_1 \overline \ell_2)$ for operators of type $\ell_1 \overline{\ell}_2 \overline{q} q$ with 
photon $\gamma (k)$ attached to the SM dipole penguin vertex.  The black circles represent the four-fermion LFV vertex (Eq. (\ref{eqn:Llq})) and the grey circles with the black border 
represent the SM dipole penguin vertex (Eq. (\ref{eqn:penguin})).}
\label{3bodydecaydiagramsCD}
\end{figure}

The scalar functions $A_i^{P\ell_1\ell_2} (p^2, ...)$ introduced in Eq.~(\ref{AtoMfinal}) can only depend on kinematical invariants and form factors. 
These functions can be calculated on the lattice or using other non-perturbative methods. 
Examining the four-fermion Lagrangian of Eq.~(\ref{eqn:Llq}) one can find that the contributions of Figs. (\ref{3bodydecaydiagramsAB}), (\ref{3bodydecaydiagramsEFGH}),
and (\ref{3bodydecaydiagramsCD}) to $A_i^{P\ell_1\ell_2}$ could be written in terms of the form factors for $P(p) \to \gamma (k)$ transitions used to parameterize lepton flavor conserving decays, such as $P^+ \to \gamma \ell^+ \bar\nu$ or $P^0 \to \gamma \ell \bar\ell$.  These form factors are defined as
\cite{Kruger:2002gf,Melikhov:2017pwu,Guadagnoli:2016erb,Kozachuk:2016ypz}
\begin{eqnarray}\label{FF}
\langle \gamma(k)|\overline q_1 \gamma^\mu \gamma_5 q_2 | P(p) \rangle &=&
i \sqrt{4\pi\alpha} \ \varepsilon^*_\alpha \!(k) \left[g^{\alpha \mu}p \cdot k - p^{\alpha} k^{\mu} \right] f_A^P[Q^2,k^2],\label{eqn:axialvectorFF} \\
\langle \gamma(k)|\overline q_1 \gamma^\mu q_2 | P(p) \rangle &=&
\sqrt{4\pi \alpha} \ \varepsilon^*_{\nu} \!(k) \epsilon^{\mu\nu\alpha\beta} p_\alpha k_\beta f_V^P[Q^2,k^2], \label{eqn:vectorFF}\\
\langle \gamma^*(k)|\overline q_1 \sigma^{\mu\nu} q_2 | P(p) \rangle &=&
i\sqrt{4\pi \alpha} \ \varepsilon^*_{\alpha} \!(k) \Bigg[\epsilon^{\mu\nu\alpha\beta} k_\beta f_{T1}^P[Q^2,k^2] + \left(p^{\alpha}-\frac{p \cdot k}{k^2} k^{\alpha} \right) 
\epsilon^{\mu \nu \rho \beta} p_{\rho} k_{\beta} f_{T2}^P[Q^2,k^2] \nonumber \\
&+& \left( \epsilon^{\mu \nu \alpha \rho} p_{\rho} + \frac{k^{\alpha}}{k^2} \epsilon^{\mu \nu \rho \beta} p_{\rho} k_{\beta} \right)f_{T3}^P[Q^2,k^2] \Bigg]. \label{eqn:offshelltensorFF}
\end{eqnarray}
Here $Q=p-k$ and the tensor form factors are defined for an off-shell photon. The tensor form factors $f_{T1,2,3}^P[k_1^2,k_2^2]$ are functions of two 
variables: $k_1$, which is the momentum flowing from a vertex associated with the tensor current, and $k_2$, which is the momentum of the photon emitted 
from the valence quark of the meson. Note that for the on-shell photon $k^2=0$, there exist a relationship between $f_{T2}^P$ and $f_{T3}^P$. Gauge invariance implies that 
$f_{T3}^P[Q^2,0] = (p \cdot k) f_{T2}^P[Q^2,0]$, so the tensor matrix element simplifies to \cite{Kruger:2002gf}
\begin{eqnarray}\label{eqn:onshelltensorFF}
\langle \gamma(k)|\overline q_1 \sigma_{\mu\nu} q_2 | P(p) \rangle &=&
i\sqrt{4\pi \alpha} \ \varepsilon^{*\alpha} \!(k) \Big[
\epsilon_{\mu\nu\alpha\beta} k^\beta f_{T1}^P[Q^2,0] \\
&& + \left(p_\alpha \epsilon_{\mu\nu\rho\beta} p^\rho k^\beta + p\cdot k \epsilon_{\mu\nu\alpha\beta} p^\beta \right) f_{T2}^P[Q^2,0] \Big]. \nonumber
\end{eqnarray}

Using Eqs. (\ref{eqn:axialvectorFF}), (\ref{eqn:vectorFF}), and (\ref{eqn:onshelltensorFF}) we can calculate the scalar function contributions of the axial, vector, and tensor 
operators from the Lagrangian in Eq.~(\ref{eqn:Llq}) of type ${\cal O} \sim (\ell_1 \overline{\ell}_2) (\overline{q}_1 q_2)$ where $q_1 \neq q_2$, which are found in Fig. (\ref{3bodydecaydiagramsAB}).  
The contributions of these diagrams to the scalar functions $A_i^{P \ell_1 \ell_2}$ are
\begin{eqnarray}  \label{eqn:A12345678ab}
	\begin{split}
		A_1^{\ref{3bodydecaydiagramsAB}ab} = \makebox[\widthof{$-$}][r]{} & 
			\makebox[\widthof{$\tfrac{i 2 \sqrt{4 \pi \alpha}}{\Lambda^2}$}][r]{$\tfrac{\sqrt{4 \pi \alpha}}{2 \Lambda^2}$} \left(C_{VR}^{q_1 q_2 \ell_1 \ell_2}
			-C_{VL}^{q_1 q_2 \ell_1 \ell_2}\right) y m_P f^P_V[m_{12}^2,0] \\
 			-  & \makebox[\widthof{$\tfrac{i 2 \sqrt{4 \pi \alpha}}{\Lambda^2}$}][r]{$\tfrac{\sqrt{4 \pi \alpha}}{\Lambda^2}$} 
 			\left(C_{TR}^{q_1 q_2 \ell_1 \ell_2}-C_{TL}^{q_1 q_2 \ell_1 \ell_2}\right)  
 			y m_P m_{H} G_F \left(f^P_{T1}[m_{12}^2,0] + \tfrac{m_P^2-m_{12}^2}{2} f^P_{T2}[m_{12}^2,0] \right) \text{,} \\
		A_3^{\ref{3bodydecaydiagramsAB}ab} = - & \makebox[\widthof{$\tfrac{i 2 \sqrt{4 \pi \alpha}}{\Lambda^2}$}][r]{$\tfrac{2 \sqrt{4 \pi \alpha}}{\Lambda^2}$}
			 \left( C_{TR}^{q_1 q_2 \ell_1 \ell_2}-C_{TL}^{q_1 q_2 \ell_1 \ell_2}\right) y m_P m_H G_F f^P_{T2}[m_{12}^2,0] \text{,} \\
		A_5^{\ref{3bodydecaydiagramsAB}ab} = - & \makebox[\widthof{$\tfrac{i 2 \sqrt{4 \pi \alpha}}{\Lambda^2}$}][r]{$\tfrac{\sqrt{4 \pi \alpha}}{2 \Lambda^2}$}
			\left( C_{AR}^{q_1 q_2 \ell_1 \ell_2} + C_{AL}^{q_1 q_2 \ell_1 \ell_2}\right) f^P_A[m_{12}^2]  \\
			+ & \makebox[\widthof{$\tfrac{i 2 \sqrt{4 \pi \alpha}}{\Lambda^2}$}][r]{$\tfrac{\sqrt{4 \pi \alpha}}{\Lambda^2}$}
			 \left( C_{TR}^{q_1 q_2 \ell_1 \ell_2}-C_{TL}^{q_1 q_2 \ell_1 \ell_2}\right) y^2 m_P^2 m_H G_F f^P_{T2}[m_{12}^2,0] \text{, and} \\
		A_7^{\ref{3bodydecaydiagramsAB}ab} = \makebox[\widthof{$-$}][r]{} & 
			\makebox[\widthof{$\tfrac{i 2 \sqrt{4 \pi \alpha}}{\Lambda^2}$}][r]{$\tfrac{\sqrt{4 \pi \alpha}}{\Lambda^2}$} 
			\left(C_{VR}^{q_1 q_2 \ell_1 \ell_2}-C_{VL}^{q_1 q_2 \ell_1 \ell_2}\right) f^P_V[m_{12}^2,0] \text{.} \\
	\end{split} 
\end{eqnarray}
Note that in this section (e.g. in writing Eq.~(\ref{eqn:A12345678ab})) we suppressed the previously used superscript of $P \ell_1 \ell_2$ in favor of a superscript related to the associated 
diagrams, which consists of the figure number and sub-figure letters (i.e. $\ref{3bodydecaydiagramsAB}ab$).  We only show the odd subscript scalar function equations.  
The even subscript equations can be found from the odd subscript equations by replacing the left-handed WCs by their negative magnitudes (i.e. $C_{VL} \to - C_{VL}$, $C_{AL} \to - C_{AL}$, etc. ) 
and multiplying the odd subscript scalar function by the imaginary constant $i$.  This may be used to find $A_2$ from $A_1$, $A_4$ from $A_3$, $A_6$ from $A_5$, and $A_8$ from $A_7$ 
and is true throughout this section.

There is contribution in Fig. (\ref{3bodydecaydiagramsAB}) from the pseudo-scalar operators of the Lagrangian in Eq.~(\ref{eqn:Llq}). This can be seen 
by taking a matrix element of the divergence of axial current to relate the axial and pseudo-scalar matrix elements,
\begin{equation}
\langle \gamma(k)|\overline q_1 \gamma_5 q_2 | P(p) \rangle 
= -\frac{1}{m_{q_1}+m_{q_2}} p^\mu \langle \gamma(k)|\overline q_1 \gamma_\mu \gamma_5 q_2 | P(p) \rangle,
\end{equation}
and using Eq.~(\ref{FF}) to get
\begin{equation}
\langle \gamma(k)|\overline q_1 \gamma_5 q_2 | P(p) \rangle = 0.
\end{equation}
A similar argument can be made to prove that the scalar operators also do not give form factor contributions.

\begin{figure}
\subfigure[]{\includegraphics[scale=0.5]{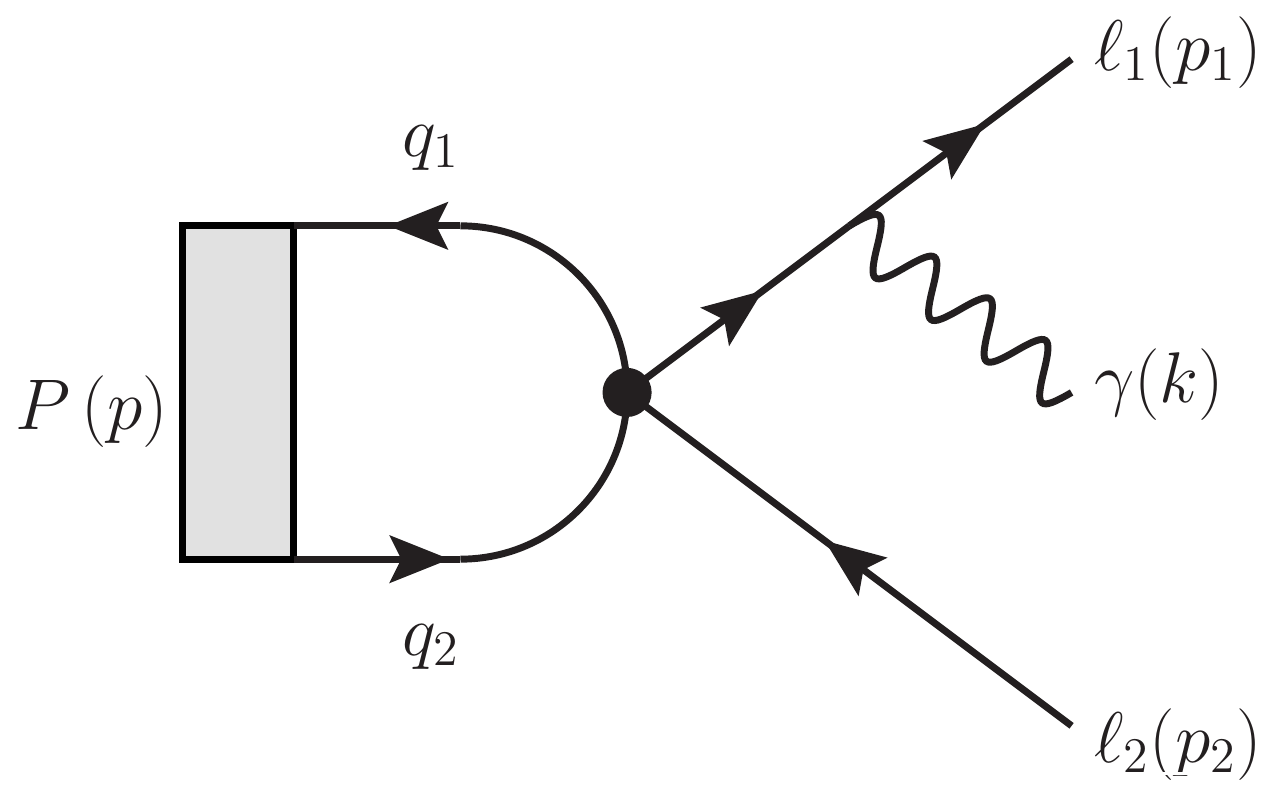} \label{diagram i}}
\hspace{0.5in}
\subfigure[]{\includegraphics[scale=0.5]{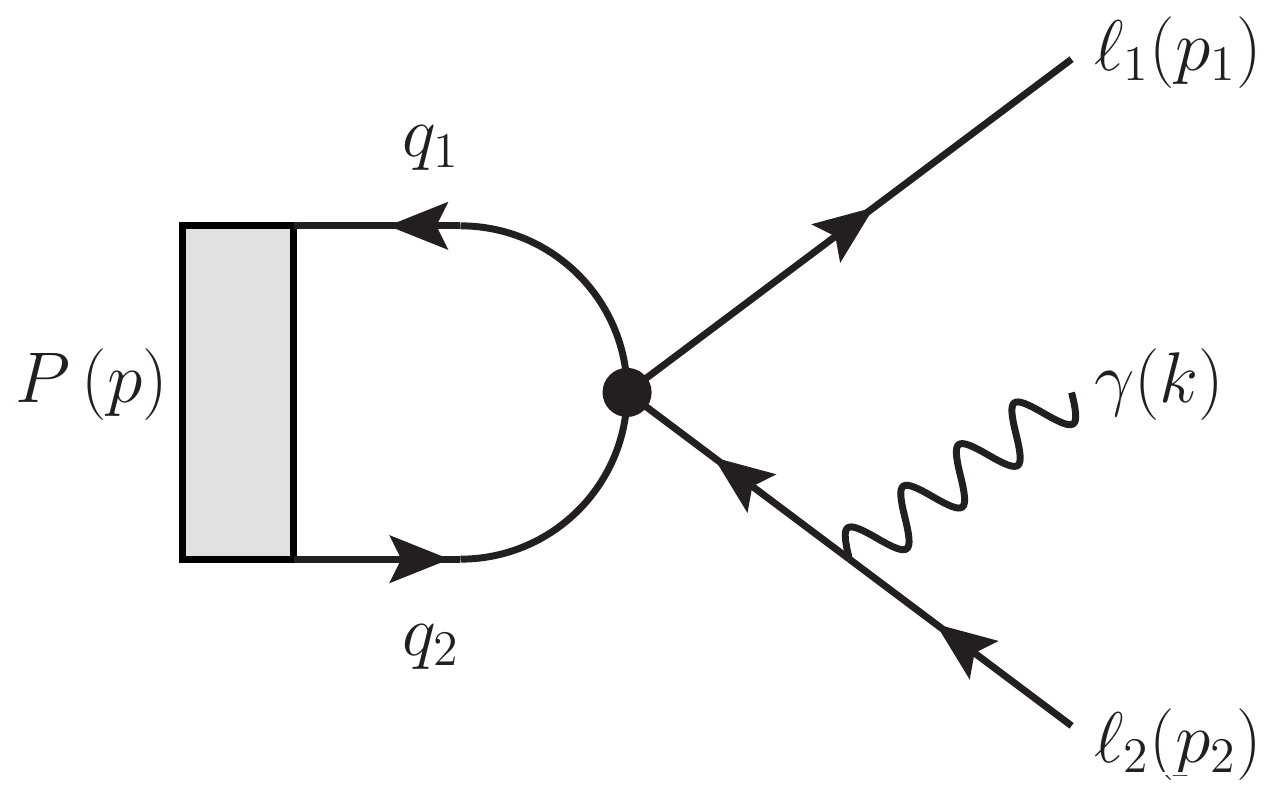} \label{diagram j}}
\caption{Bremsstrahlung diagrams for ${\cal A} (P \to \gamma \ell_1 \overline \ell_2)$ for operators of type ${\cal O} \sim (\ell_1 \overline{\ell}_2) (\overline{q}_1 q_2)$ where $q_1 \neq q_2$.  The black circles represent the four-fermion LFV vertex defined in ${\cal L}_{eff}$ of Eq. (\ref{eqn:Llq}).}
\label{3bodydecaydiagramsIJ}
\end{figure}

The bremsstrahlung diagrams in Fig.~(\ref{3bodydecaydiagramsIJ}) are calculated similarly to the two-body decays of Sect. \ref{Spin0LLgam} using the matrix element of Eq.~(\ref{DeConP}). 
We have given the photon a small mass, $m_{\gamma}$, to regulate the infrared divergences.  This divergence only appears in the quark flavor changing axial and pseudoscalar 
operator terms of the scalar functions, Eq.~(\ref{eqn:A1234ij}), so the photon mass is set to zero for the non-divergent terms. The same is true for the differential decay rate in 
Eq.~(\ref{eqn:DiffDecayRate}). The axial and pseudoscalar operator scalar function terms are defined here as
\begin{align}
	\begin{split} \label{eqn:A1234ij}
		A_1^{\ref{3bodydecaydiagramsIJ}ab} = \makebox[\widthof{$\tfrac{i 2 \sqrt{4 \pi \alpha}}{\Lambda^2}$}][r]{$\tfrac{\sqrt{4 \pi \alpha}}{2 \Lambda^2}$}
			& \left(C_{AR}^{q_1 q_2 \ell_1 \ell_2} \! +C_{AL}^{q_1 q_2 \ell_1 \ell_2} \! + m_P^2 G_F  
			\left(C_{PR}^{q_1 q_2 \ell_1 \ell_2} \! +C_{PL}^{q_1 q_2 \ell_1 \ell_2}\right) \! \right) 
			\! \tfrac{y m_P f_P \left(m_P^2+m^2_\gamma-m_{12}^2\right)}{m_{13}^2\left(m_{23}^2-m_P^2y^2\right)} \text{,} \\
		A_3^{\ref{3bodydecaydiagramsIJ}ab} =  \makebox[\widthof{$\tfrac{i 2 \sqrt{4 \pi \alpha}}{\Lambda^2}$}][r]{$\tfrac{2 \sqrt{4 \pi \alpha}}{\Lambda^2}$}
			& \left(C_{AR}^{q_1 q_2 \ell_1 \ell_2} \! -C_{AL}^{q_1 q_2 \ell_1 \ell_2} \! + m_P^2 G_F  
			\left(C_{PR}^{q_1 q_2 \ell_1 \ell_2} \! -C_{PL}^{q_1 q_2 \ell_1 \ell_2}\right) \! \right)
			\! \tfrac{y m_P f_P}{m_{13}^2\left(m_{23}^2-m_P^2y^2\right)} \text{.}
	\end{split}
\end{align}

The dipole operator diagrams of Eq.~(\ref{eqn:LD}) found in Fig. (\ref{3bodydecaydiagramsEFGH}) contain contributions from the SM dipole penguin operator, Eq.~(\ref{eqn:penguin}).  
This is directly related to both the on and off-shell tensor matrix elements in Eqs.~(\ref{eqn:offshelltensorFF}) and (\ref{eqn:onshelltensorFF}) from which we need to 
write matrix elements of the form $\langle \gamma(k)|\overline q_1 \sigma^{\mu\nu} (1 \pm \gamma_5) q_2 | P(p) \rangle$.  These can be found by using the relation 
$\sigma_{\mu \nu} \gamma_5 = - \tfrac{i}{2} \epsilon_{\mu \nu \alpha \beta} \sigma^{\alpha \beta}$, which yields:
\begin{align}
\begin{split}
\langle \gamma(k)|\overline q_1 \sigma^{\mu\nu} (1 \pm \gamma_5) q_2 | P(p) \rangle Q_{\nu} &= i \sqrt{4 \pi \alpha} \ \varepsilon^{*}_{\alpha}\!(k) \big\lbrace \left(f_{T1}^P[Q^2,0] + p \cdot k f_{T2}^P[Q^2,0] \right) \epsilon^{p k \alpha \mu} \label{eqn:onShellPenguinMatrixElement} \\
& \ \ \pm i \left(f_{T1}^P[Q^2,0] + p \cdot Q f_{T2}^P[Q^2,0] \right) \left( g^{\alpha \mu} p \cdot k - p^{\alpha} k^{\mu} \right) \big\rbrace \text{,}
\end{split} \\
\begin{split}
\langle \gamma^*(Q)|\overline q_1 \sigma^{\mu\nu} (1 \pm \gamma_5) q_2 | P(p) \rangle k_{\nu} &=f i \sqrt{4 \pi \alpha} \ \varepsilon^{*}_{\alpha}\!(Q)\left\lbrace \epsilon^{p k \mu \alpha} \pm i \left( g^{\alpha \mu} p \cdot k - p^{\mu} k^{\alpha} \right)  \right\rbrace \label{eqn:offShellPenguinMatrixElement} \\
& \hspace{2.7cm} \times \left( f_{T1}^P[0,Q^2] + f_{T3}^P[0,Q^2] \right) \text{.}
\end{split}
\end{align}
The on-shell matrix element in Eq.~(\ref{eqn:onShellPenguinMatrixElement}) contributes to Figs. \ref{diagram e} and \ref{diagram f}. While the off-shell matrix element in 
Eq.~(\ref{eqn:offShellPenguinMatrixElement}) is necessary for calculating the dipole operator contributions of the diagrams in Figs. \ref{diagram g} and \ref{diagram h}.  
In these diagrams, the lepton current is attached to the photon coming from the meson's valence quarks and so $Q \leftrightarrow k$ when we calculate 
Eq.~(\ref{eqn:offShellPenguinMatrixElement}). Using these matrix elements we find the dipole operator components of the scalar functions which are
\begin{align}
	\begin{split} \label{eqn:A123456efgh}
		A_1^{\ref{3bodydecaydiagramsEFGH}abcd} = - & \tfrac{1}{\Lambda^2} \left( C_{DR}^{\ell_1 \ell_2} - C_{DL}^{\ell_1 \ell_2} \right) \tfrac{4 \pi \alpha}{\pi^2} 
			\ y m_P m_H \tfrac{G_F}{\sqrt{2}} C_{7 \gamma} \sum_q \lambda_q \ f^P_{T, I} \text{,} \\
		A_3^{\ref{3bodydecaydiagramsEFGH}abcd} = \makebox[\widthof{$-$}][r]{}
			& \tfrac{2}{\Lambda^2} \tfrac{4 \pi \alpha}{\pi^2} 
			\makebox[\widthof{$\tfrac{y^2 m_P^2 m_H}{m_{12}^2}$}][c]{$\tfrac{y m_P m_H}{m_{12}^2}$} \tfrac{G_F}{\sqrt{2}} C_{7 \gamma} \sum_q \lambda_q 
			\left(\left( C_{DR}^{\ell_1 \ell_2} - C_{DL}^{\ell_1 \ell_2} \right) f^P_{T, \text{I}} - 
			\left( C_{DR}^{\ell_1 \ell_2} + C_{DL}^{\ell_1 \ell_2} \right) f^P_{T, \text{II}}\right) \text{,} \\
		A_5^{\ref{3bodydecaydiagramsEFGH}abcd} = - & \tfrac{1}{\Lambda^2} \tfrac{4 \pi \alpha}{\pi^2} \tfrac{y^2 m_P^2 m_H}{m_{12}^2} \tfrac{G_F}{\sqrt{2}} 
			C_{7 \gamma} \sum_q \lambda_q \left(\left( C_{DR}^{\ell_1 \ell_2} - C_{DL}^{\ell_1 \ell_2} \right) f^P_{T, \text{I}} - 
			\left( C_{DR}^{\ell_1 \ell_2} + C_{DL}^{\ell_1 \ell_2} \right) f^P_{T, \text{II}} \right) \text{,}
	\end{split}
\end{align}
where we have used the shorthand notations $f^P_{T, \text{I}}$ and $f^P_{T, \text{II}}$ that we define as 
\begin{align}
	\begin{split} \label{eqn:dipoleFTI&II}
		f^P_{T, \text{I}} = & f_{T1}^P[m_{12}^2,0]+f_{T1}^P[0,m_{12}^2] 
			+ \tfrac{m_P^2-m_{12}^2}{2} f_{T2}^P[m_{12}^2,0] + f_{T3}^P[0,m_{12}^2] \text{ and}\\
		f^P_{T, \text{II}} = & f_{T1}^P[m_{12}^2,0]+f_{T1}^P[0,m_{12}^2] 
			+ \tfrac{m_P^2+m_{12}^2}{2} f_{T2}^P[m_{12}^2,0] + f_{T3}^P[0,m_{12}^2] \text{.}
	\end{split}
\end{align}


So far we have not addressed the contributions of the diagrams in Fig. (\ref{3bodydecaydiagramsCD}). These diagrams contain 
contributions from the axial, vector, and tensor operators from the Lagrangian in Eq.~(\ref{eqn:Llq}) of type $\ell_1 \overline{\ell}_2 \overline{q} q$, where the 
quarks are both the same flavor.  As was the case for the four-fermion operators that had a flavor change on both the quark side and lepton side, the scalar and 
pseudo-scalar operators do not contribute.  
We can calculate the contributions of the vector operators using the same tensor matrix element as in Eq.~(\ref{eqn:offShellPenguinMatrixElement}), but 
with one important modification.  The form factors are the sum of two form factors related to each quark flavor, $f_{Ti} = \tilde{f}_{Ti}^{q_1} + \tilde{f}_{Ti}^{q_2}$ 
(e.g. see \cite{Beyer:2001zn}).  For convenience we will use a definition with the quark charge explicitly included in the formula, 
$f_{Ti} = Q_{q_1} f_{Ti}^{q_1} +  Q_{q_2} f_{Ti}^{q_2}$.  This is important because in the case of Fig. \ref{diagram c} we only have contributions from 
$f_{Ti}^{q_1}$ and in Fig. \ref{diagram d} we only have  $f_{Ti}^{q_2}$.
\begin{align}
	\begin{split} \label{eqn:A125678cd}
		A_1^{\ref{3bodydecaydiagramsCD}ab} = - & \tfrac{\sqrt{4 \pi \alpha}}{\pi^2 \Lambda^2} 
			\sum_{j = 1}^2  \left(C_{VR}^{q_j \ell_1 \ell_2}-C_{VL}^{q_j \ell_1 \ell_2}\right) 
			\! \tfrac{y m_P}{2} \tfrac{G_F}{\sqrt{2}} C_{7 \gamma} \sum_q \lambda_q^P \left(f^{P,q_j}_{T1}[0,m_{12}^2] + f^{P,q_j}_{T3}[0,m_{12}^2] \right) \text{,} \\
		A_5^{\ref{3bodydecaydiagramsCD}ab} = \makebox[\widthof{$-$}][c]{} &\tfrac{\sqrt{4 \pi \alpha}}{\pi^2 \Lambda^2}
			\sum_{j = 1}^2  \left(C_{VR}^{q_j \ell_1 \ell_2}+C_{VL}^{q_j \ell_1 \ell_2}\right) 
			\! \makebox[\widthof{$\tfrac{y m_P}{2}$}][c]{$\tfrac{m_H}{2}$} \tfrac{G_F}{\sqrt{2}} C_{7 \gamma} \sum_q \lambda_q^P \left(f^{P,q_j}_{T1}[0,m_{12}^2] 
			+ f^{P,q_j}_{T3}[0,m_{12}^2] \right) \text{,} \\
		A_7^{\ref{3bodydecaydiagramsCD}ab} = - & \tfrac{\sqrt{4 \pi \alpha}}{\pi^2 \Lambda^2}
			\sum_{j = 1}^2  \left(C_{VR}^{q_j \ell_1 \ell_2}-C_{VL}^{q_j \ell_1 \ell_2}\right) 
			\! \makebox[\widthof{$\tfrac{y m_P}{2}$}][c]{$m_H$} \tfrac{G_F}{\sqrt{2}} C_{7 \gamma} \sum_q \lambda_q^P \left(f^{P,q_j}_{T1}[0,m_{12}^2] 
			+ f^{P,q_j}_{T3}[0,m_{12}^2] \right) \text{.}
	\end{split}
\end{align}
%

Applying this information to the decays of $B^0_q$, $\bar{D}^0$, and $K^0$ mesons shown in Figs. (\ref{3bodydecaydiagramsAB})--(\ref{3bodydecaydiagramsCD}), we find 
that each scalar function $A_i^{P\ell_1\ell_2}$ is written as
\begin{equation} \label{eqn:ScalarFunction}
A_i^{P\ell_1\ell_2} (p^2, ...) = A_i^{\ref{3bodydecaydiagramsAB}ab} + A_i^{\ref{3bodydecaydiagramsIJ}ab} + A_i^{\ref{3bodydecaydiagramsEFGH}abcd} + 
A_i^{\ref{3bodydecaydiagramsCD}ab} \quad \left(i = 1 \text{--} 8\right) \text{,}
\end{equation}
which are functions of model independent form factors and decay constants.

\section{Results}\label{Results}

Unfortunately, no experimental limits on the branching ratios of radiative lepton-flavor violating decays exist to constrain all of the applicable Wilson 
coefficients of the effective Lagrangian of Eq.~(\ref{eqn:Leff}). We encourage our colleagues from the LHC and KEK to study these decays. 
However, some information about Wilson coefficients is available from other transitions, such as two-body decays discussed in Sect. \ref{Spin0LLgam}.
In this section we use this information, along with the assumption of single operator dominance to derive the expectations for the size of 
the radiative LFV decays, if driven by those operators. These upper limits are presented in Tables \ref{tab:BdecaylimitsFF} and \ref{tab:DKdecaylimitsFF} and the differential decay 
rates are plotted in Figs. (\ref{fig:vectorq1q2l1l2decayplots})--(\ref{fig:pseudoscalarq1q2l1l2decayplots}) of Section \ref{OurResults}.  

All of the form factors and numerical constants, unless previously mentioned, used to obtain the results in this section may be found in 
Appendix \ref{FormFactors}.  In some cases where form factors are currently unknown, we apply a 
constituent quark model to estimate the relevant contribution.  The quark model approach and results may be found in Appendix \ref{QuarkModel}.

\subsection{Spectra}\label{OurResults}

Inputting the scalar functions of Eq. (\ref{eqn:ScalarFunction}) in the differential decay rate, Eq. (\ref{eqn:DiffDecayRate}), and integrating over the Mandelstam variables $m_{23}^2$ and $m_{12}^2$, we calculate the differential decay rate, $d \Gamma/d m_{12}^2$, and total decay rate, $\Gamma\left(P \to \gamma \ell_1 \overline \ell_2\right)$.  Using these results we may predict the differential decay spectra for individual operators, $\left(1/\Gamma \right) \left(d \Gamma/d E_{\gamma} \right)$.  Where we make the variable change from $m_{12}^2$ to $E_{\gamma}$, the photon energy in the meson rest frame, and normalize to the total decay rate.  This analysis requires the practical assumption of single operator dominance so that the unknown WCs of individual operators will cancel between the differential and total decay rates.

\begin{figure} 
\subfigure[]{\includegraphics[scale=0.5]{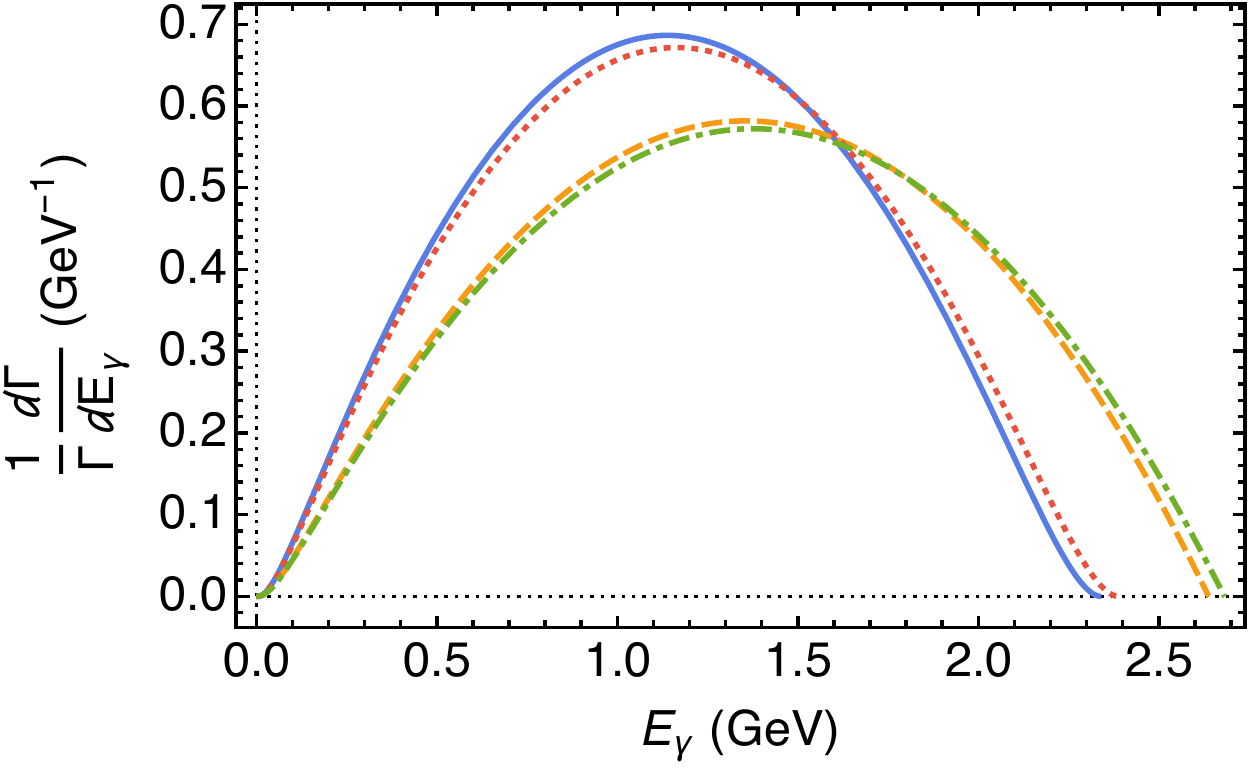} \label{fig:bd(s)l1l2vectordecayplots}}
\hspace{0.5in}
\subfigure[]{\includegraphics[scale=0.5]{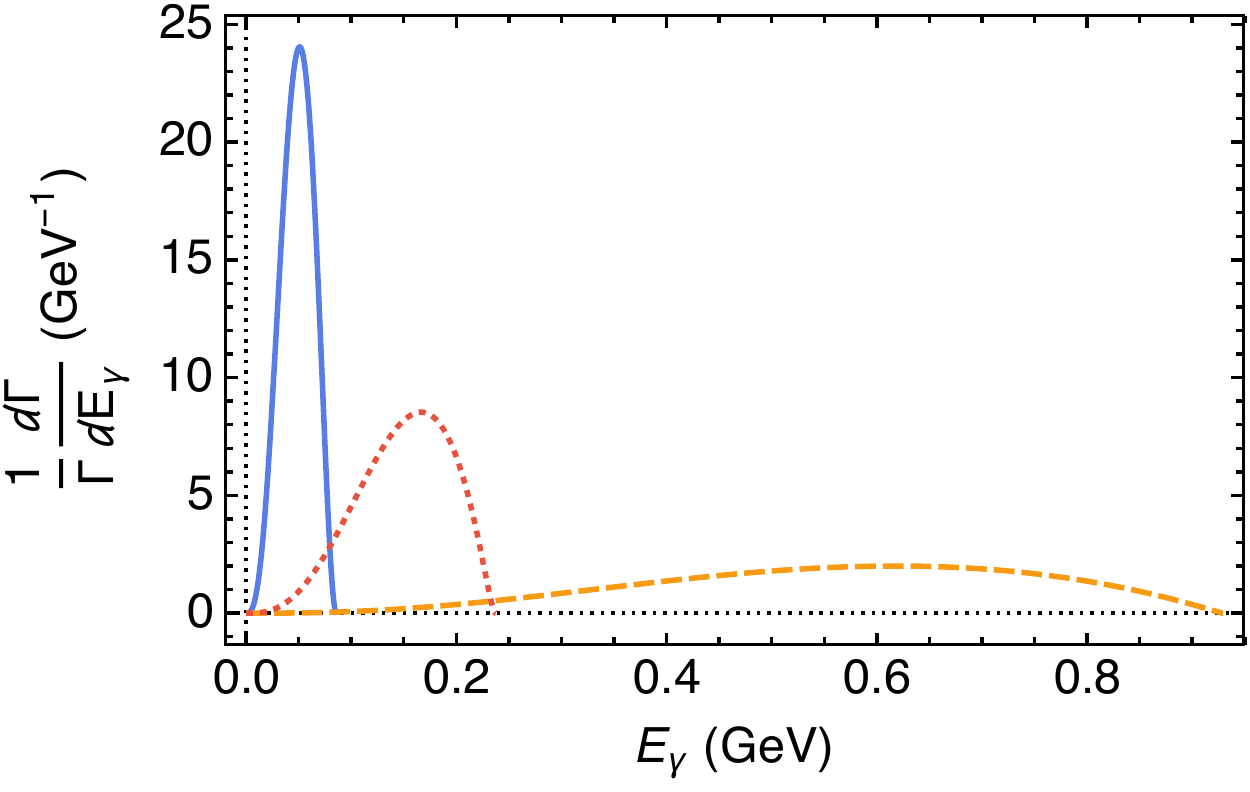} \label{fig:cu(ds)l1l2vectordecayplots}}
\caption{\label{fig:vectorq1q2l1l2decayplots} Vector operator (${\cal O} \sim (\ell_1 \overline{\ell}_2) (\overline{q}_1 q_2)$ where $q_1 \neq q_2$) differential decay plots as functions of photon energy $E_{\gamma}$: (a) $B_d \to \gamma \mu \tau$ or $\gamma e \tau$ (solid blue curve), $B_d \to \gamma e \mu$ (short-dashed gold curve), $B_s \to \gamma \mu \tau$ or $\gamma e \tau$ (dotted red curve), $B_s \to \gamma e \mu$ (dot-dashed green curve); (b) $D \to \gamma e \tau$ (solid blue curve),  $D \to \gamma e \mu$ (short-dashed gold curve), $K \to \gamma e \mu$ (dotted red curve)}
\end{figure}

\begin{figure}
\subfigure[]{\includegraphics[scale=0.5]{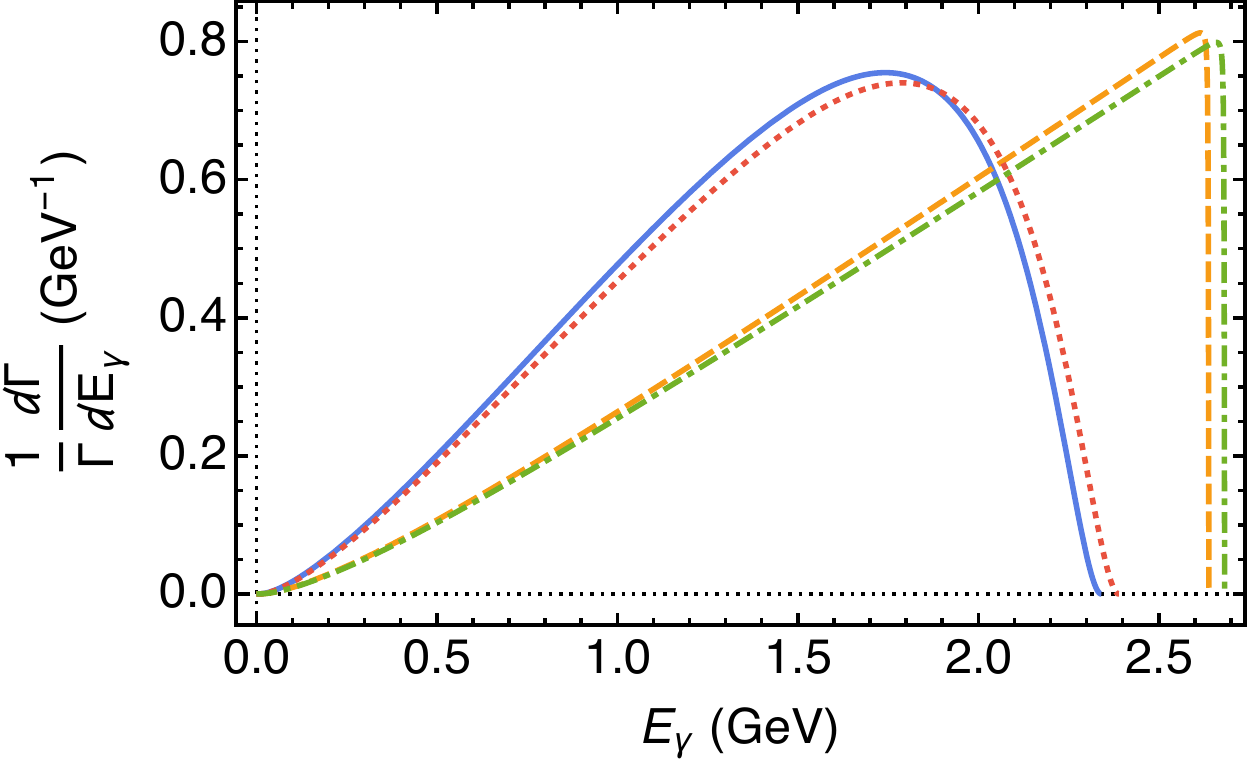} \label{fig:bd(s)l1l2vectordecayplots}}
\hspace{0.5in}
\subfigure[]{\includegraphics[scale=0.5]{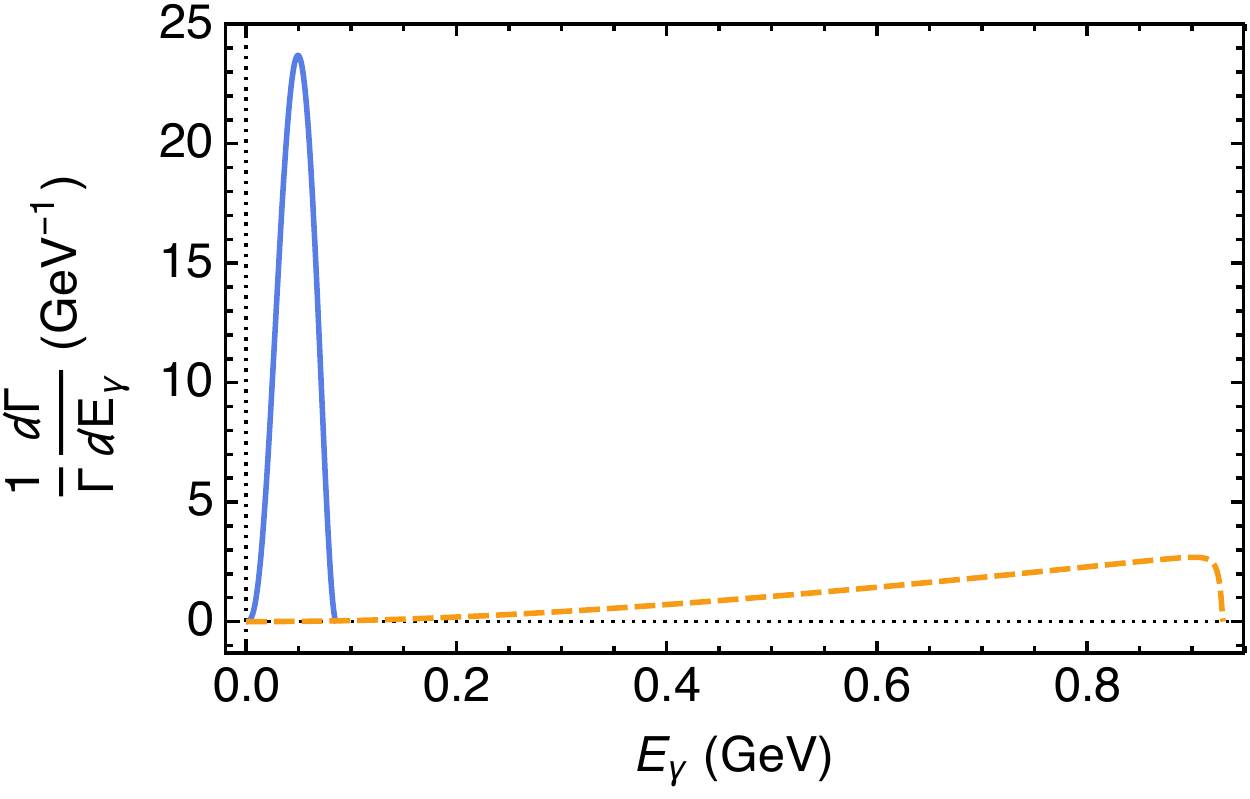} \label{fig:cu(ds)l1l2vectordecayplots}}
\caption{\label{fig:tensorq1q2l1l2decayplots} Tensor operator (${\cal O} \sim (\ell_1 \overline{\ell}_2) (\overline{q}_1 q_2)$ where $q_1 \neq q_2$) differential decay plots as functions of photon energy $E_{\gamma}$: (a) $B_d \to \gamma \mu \tau$ or $\gamma e \tau$ (solid blue curve), $B_d \to \gamma e \mu$ (short-dashed gold curve), $B_s \to \gamma \mu \tau$ or $\gamma e \tau$ (dotted red curve), $B_s \to \gamma e \mu$ (dot-dashed green curve); (b) $D \to \gamma e \tau$ (solid blue curve),  $D \to \gamma e \mu$ (short-dashed gold curve)}
\end{figure}

\begin{figure} 
\subfigure[]{\includegraphics[scale=0.5]{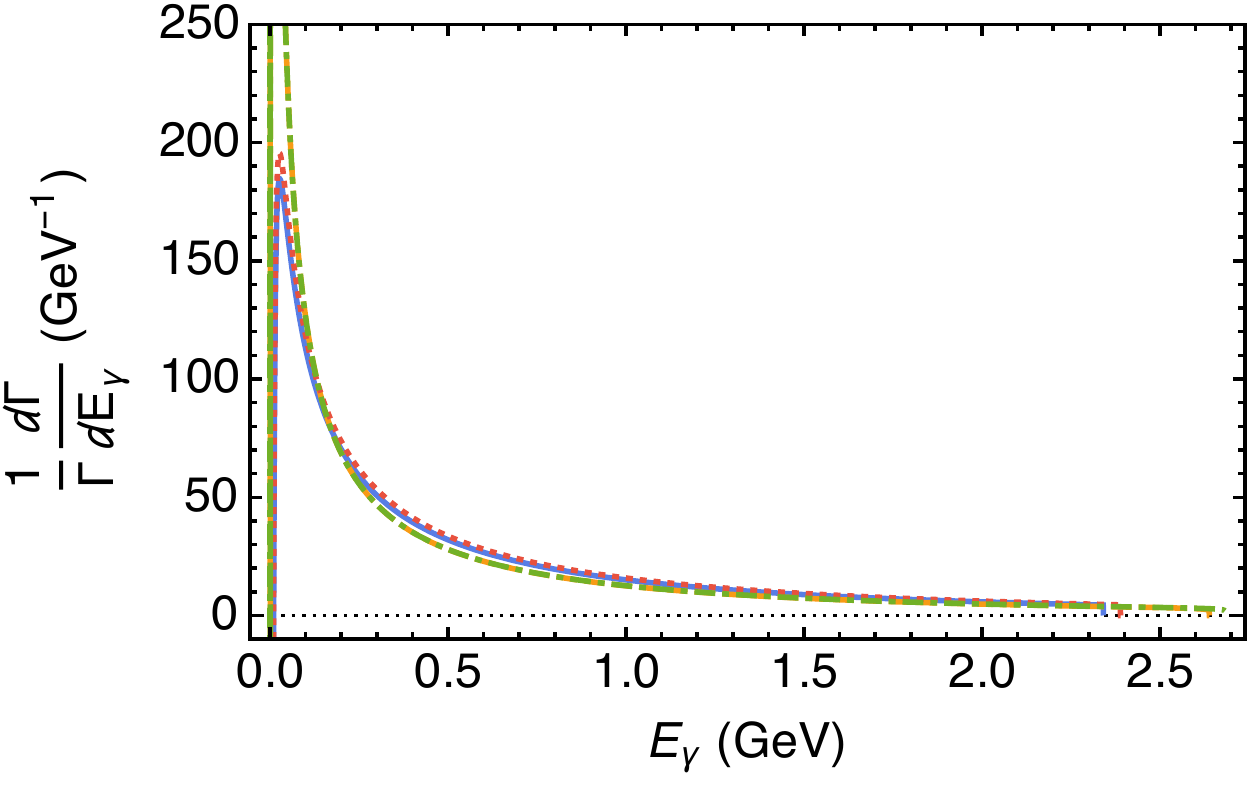} \label{fig:bd(s)l1l2axialdecayplots}}
\hspace{0.5in}
\subfigure[]{\includegraphics[scale=0.5]{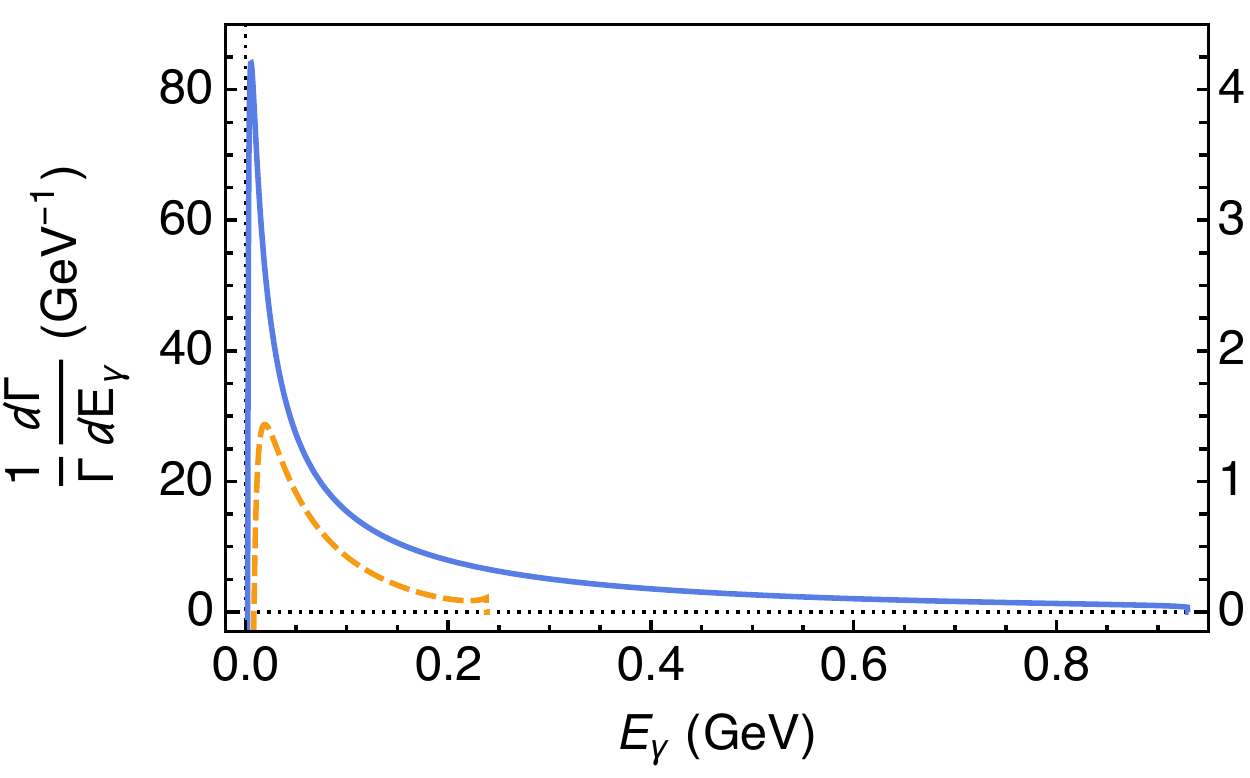} \label{fig:cu(ds)l1l2axialdecayplots}}
\caption{\label{fig:axialq1q2l1l2decayplots} Axial operator (${\cal O} \sim (\ell_1 \overline{\ell}_2) (\overline{q}_1 q_2)$ where $q_1 \neq q_2$) differential decay plots as functions of photon energy $E_{\gamma}$: (a) $B_d \to \gamma \mu \tau$ or $\gamma e \tau$ (solid blue curve), $B_d \to \gamma e \mu$ (short-dashed gold curve), $B_s \to \gamma \mu \tau$ or $\gamma e \tau$ (dotted red curve), $B_s \to \gamma e \mu$ (dot-dashed green curve); (b) left scale $D \to \gamma e \mu$ (solid blue curve),  right scale $K \to \gamma e \mu$ (short-dashed gold curve)}
\end{figure}

\begin{figure}
\subfigure[]{\includegraphics[scale=0.5]{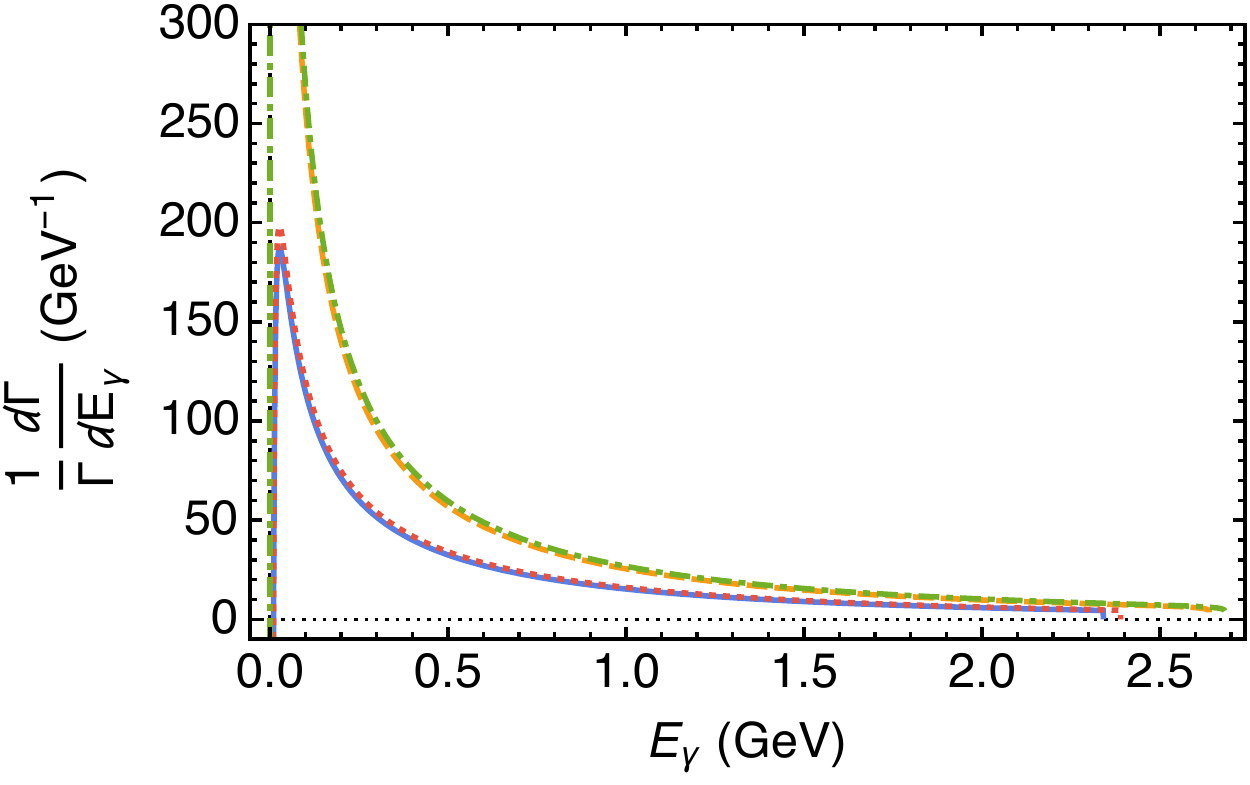} \label{fig:bd(s)l1l2pseudoscalardecayplots}}
\hspace{0.5in}
\subfigure[]{\includegraphics[scale=0.5]{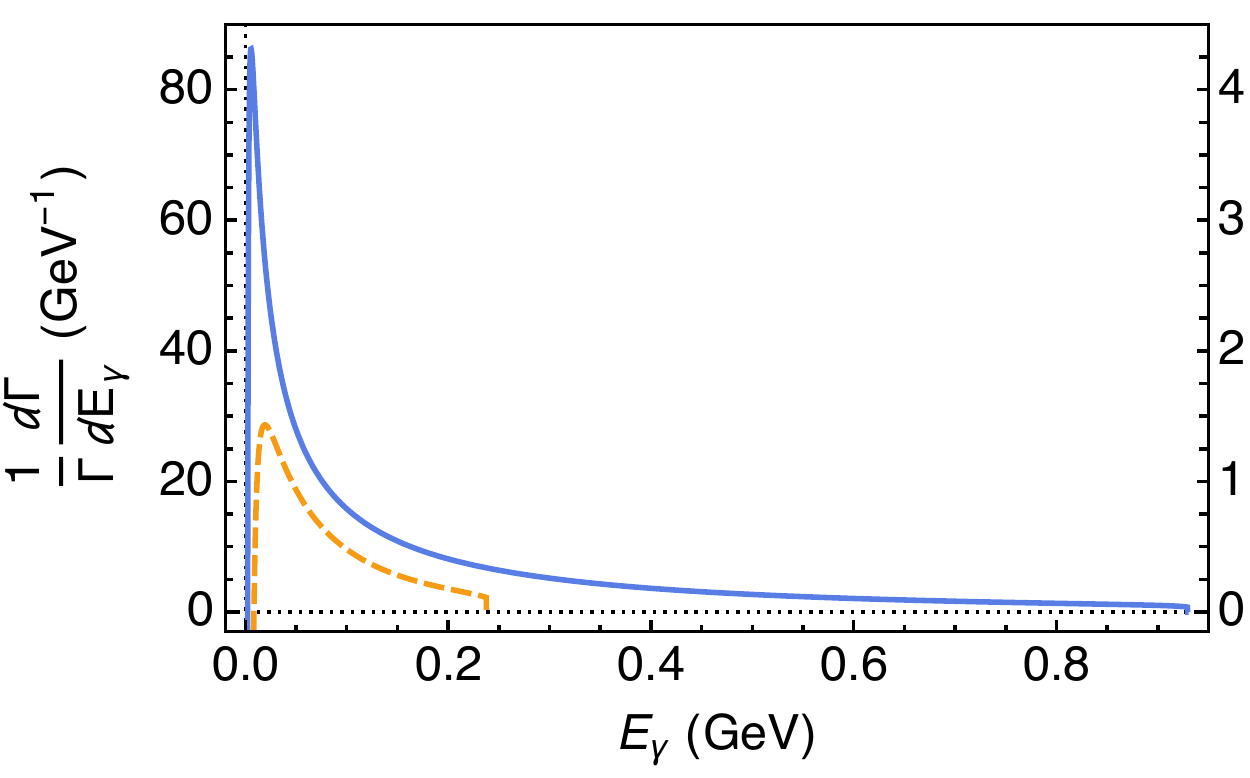} \label{fig:cu(ds)l1l2pseudoscalardecayplots}}
\caption{\label{fig:pseudoscalarq1q2l1l2decayplots} Pseudoscalar operator (${\cal O} \sim (\ell_1 \overline{\ell}_2) (\overline{q}_1 q_2)$ where $q_1 \neq q_2$) differential decay plots as functions of photon energy $E_{\gamma}$: (a) $B_d \to \gamma \mu \tau$ or $\gamma e \tau$ (solid blue curve), $B_d \to \gamma e \mu$ (short-dashed gold curve), $B_s \to \gamma \mu \tau$ or $\gamma e \tau$ (dotted red curve), $B_s \to \gamma e \mu$ (dot-dashed green curve); (b) left scale $D \to \gamma e \mu$ (solid blue curve),  right scale $K \to \gamma e \mu$ (short-dashed gold curve)}
\end{figure}

The differential decay rates for the vector and tensor operators of type ${\cal O} \sim (\ell_1 \overline{\ell}_2) (\overline{q}_1 q_2)$ where $q_1 \neq q_2$ are 

\begin{align}
		\frac{d \Gamma^{q_1 q_2 \ell_1 \ell_2}_V}{d m_{12}^2}&=\frac{C_{VR}^2+C_{VL}^2}{\Lambda^4} \frac{4 \pi \alpha}{(2 \pi)^3} \label{eqn:diffvec}
			\frac{\makebox[\widthof{$y^2 m_{q_{\text{H}}}^2 G_F^2$}][c]{$1$}}{576 m_P^2} \left(m_P^2 - m_{12}^2 \right)^3 \left( 2 m_{12}^2 - 3 m_P^2 y^2 \right) f^P_V[m_{12}^2,0] \text{,} \\
		\begin{split}
			\frac{d \Gamma^{q_1 q_2 \ell_1 \ell_2}_T}{d m_{12}^2}&=\frac{C_{TR}^2+C_{TL}^2}{\Lambda^4} \frac{4 \pi \alpha}{(2 \pi)^3} \frac{y^2 m_{q_{\text{H}}}^2 G_F^2}{288 m_P^2} 
				 \left(m_P^2 - m_{12}^2 \right)^3 \\ 
				 & \hspace{2.5cm}\times \left( \left(2 f_{T1}^P[m_{12}^2,0] + m_P^2 f_{T2}^P[m_{12}^2,0] \right)^2 + m_{12}^2 \left(f_{T2}^P[m_{12}^2,0] \right)^2 \right) \text{.} \label{eqn:difftensor}
		\end{split}
\end{align}

Here we have suppressed the superscripts of the WCs for brevity (e.g. $C_{VR}^{q_1 q_2 \ell_1 \ell_2} \to C_{VR}$).  We drop terms higher in order than $y^2$, which is a good approximation in most cases as the ratio $y$ is small.  The vector and tensor operators with flavor change on both the quark and lepton side are of particular importance to our analysis.  They cannot be constrained via two-body decays and so the three-body decay channels present us with a unique opportunity to place limits on the associated WCs.  The vector operators also have an advantage over the tensor operators because they are not chirally suppressed by quark and lepton masses.  Assuming WCs are of similar size, this means the vector operators would give a larger contribution to the overall decay rate and conversely are better constrained by experimental limits.  The differential spectra given in Eqs. (\ref{eqn:diffvec})--(\ref{eqn:difftensor}) are shown in Figs.  (\ref{fig:vectorq1q2l1l2decayplots})--(\ref{fig:tensorq1q2l1l2decayplots}).

The three-body decays considered here also provide complementary access to the axial and pseudo-scalar operators of type ${\cal O} \sim (\ell_1 \overline{\ell}_2) (\overline{q}_1 q_2)$ where $q_1 \neq q_2$.  We do not provide the equations for the individual differential decay rates as they are more cumbersome than their vector and tensor counterparts and they are better constrained via two-body decays.  Their differential spectra are plotted in Figs. (\ref{fig:axialq1q2l1l2decayplots})--(\ref{fig:pseudoscalarq1q2l1l2decayplots})  We demonstrate how well constrained these and other operators are in Sect. \ref{Limits} and Appendix \ref{ResultsQM}.


\subsection{Limits}\label{Limits}

Using the available limits on Wilson coefficients from Section \ref{Spin0LLgam} with the form factors of Appendix \ref{FormFactors}, we predict the upper threshold experiments must reach to potentially see LFV in the $P \to \gamma \ell_1 \overline{\ell}_2$ decays involving the axial and pseudo-scalar operators of type ${\cal O} \sim (\ell_1 \overline{\ell}_2) (\overline{q}_1 q_2)$ where $q_1 \neq q_2$ and dipole operators.  These upper bounds are presented in Table \ref{tab:BdecaylimitsFF} for $B^0_q$ decays and in Table \ref{tab:DKdecaylimitsFF} for $\bar{D}^0$ and $K^0_L$ decays.  $K^0_L$ is used in lieu of $K^0$ for the limits on the branching ratios due to a lack of experimental information on the total decay rate of $K^0$.  The normalized differential decay plots of $K^0$ are the same as $K^0_L$ because the normalization to the total decay rate cancels out the numerical differences (i.e. a factor of $1/\sqrt{2}$).  

\begin{table*}
\caption{\label{tab:BdecaylimitsFF} Upper limits on $B_{q}^0 \to \gamma \ell_1 \overline{\ell}_2$ branching ratios from known Wilson coefficient constraints using form factors for four-fermion axial and pseudo-scalar operators of type ${\cal O} \sim (\ell_1 \overline{\ell}_2) (\overline{q}_1 q_2)$ where $q_1 \neq q_2$.}
\begin{ruledtabular}
\begin{tabular}{lcccc} Wilson & \multicolumn{4}{c}{Upper limits} \\
coefficient & ${\cal B}(B^0_d \to \gamma \mu \tau)$ & ${\cal B}(B^0_d \to \gamma e \tau)$ & ${\cal B}(B^0_d \to \gamma e \mu)$ & ${\cal B}(B^0_s \to \gamma e \mu)$ \\
\hline
$C_{AR}^{q b \ell_1 \ell_2}$ & $9.2 \times 10^{-7}$ & $1.2 \times 10^{-6}$ & $6.5 \times 10^{-11}$ & $3.7 \times 10^{-10}$ \\
$C_{AL}^{q b \ell_1 \ell_2}$ & $9.2 \times 10^{-7}$ & $1.2 \times 10^{-6}$ & $6.5 \times 10^{-11}$ & $3.7 \times 10^{-10}$ \\
$C_{PR}^{q b \ell_1 \ell_2}$ & $9.0 \times 10^{-7}$ & $1.2 \times 10^{-6}$ & $3.2 \times 10^{-11}$ & $1.7 \times 10^{-10}$ \\
$C_{PL}^{q b \ell_1 \ell_2}$ & $9.0 \times 10^{-7}$ & $1.2 \times 10^{-6}$ & $3.2 \times 10^{-11}$ & $1.7 \times 10^{-10}$ \\
\end{tabular}
\end{ruledtabular}
\end{table*}

\begin{table*}
\caption{\label{tab:DKdecaylimitsFF} Upper limits on $\bar{D}^0 \left(u \bar c \right)$, $K^0_L \left( \left(d \overline s - s \overline d \right)/\sqrt{2}\right)  \to \gamma \ell_1 \overline{\ell}_2$ branching ratios from known Wilson coefficient constraints using form factors for four-fermion axial and pseudo-scalar operators of type ${\cal O} \sim (\ell_1 \overline{\ell}_2) (\overline{q}_1 q_2)$ where $q_1 \neq q_2$.  Note the $K^0_L$ results are for short distance (SD) interactions. }
\begin{ruledtabular}
\begin{tabular}{lcc} Wilson & \multicolumn{2}{c}{Upper limits} \\
coefficient & ${\cal B}(\bar{D}^0 \to \gamma e \mu)$ & ${\cal B}(K^0_L \to \gamma e \mu)_{SD}$ \\
\hline
$C_{AR}^{q_1 q_2 \ell_1 \ell_2}$ & $2.2 \times 10^{-10}$ & $2.3 \times 10^{-14}$ \\
$C_{AL}^{q_1 q_2 \ell_1 \ell_2}$ & $2.2 \times 10^{-10}$ & $2.3 \times 10^{-14}$ \\
$C_{PR}^{q_1 q_2 \ell_1 \ell_2}$ & $4.5 \times 10^{-9}$ & $2.2 \times 10^{-14}$ \\
$C_{PL}^{q_1 q_2 \ell_1 \ell_2}$ & $4.5 \times 10^{-9}$ & $2.2 \times 10^{-14}$ \\
\end{tabular}
\end{ruledtabular}
\end{table*}

\begin{table*}
\caption{\label{tab:DipoledecaylimitsFF} Upper limits on $B^0_q \left(q \overline{b}\right)$, $\bar{D}^0 \left(u \bar c \right) \to \gamma \ell_1 \overline{\ell}_2$ branching ratios from known dipole Wilson coefficient constraints using form factors for dipole operators. FPS stands for ``forbidden phase space.'' }
\begin{ruledtabular}
\begin{tabular}{l|ll|ccc} Leptons \hspace{0.3cm} & \multicolumn{2}{l|}{Wilson coefficient \cite{Hazard:2016fnc}} & \multicolumn{3}{c}{Predicted upper limits} \\
$\ell_1 \ell_2$ & \multicolumn{2}{l|}{(GeV$^{-2}$)} & ${\cal B}(B_d^0 \to \gamma \ell_1 \overline{\ell}_2)$ & ${\cal B}(B_s^0 \to \gamma \ell_1 \overline{\ell}_2)$ & ${\cal B}(\bar{D}^0 \to \gamma \ell_1 \overline{\ell}_2)$ \\
\hline
$\mu \tau$ & $|C_{DR}^{\ell_1 \ell_2}/\Lambda^2| = $ & $2.6 \times 10^{-10}$ \hspace{0.3cm} & $3.1 \times 10^{-28}$ & $1.2 \times 10^{-26}$ & FPS \\
$e \tau$ & & $2.7 \times 10^{-10}$ & $3.3 \times 10^{-28}$ & $1.3 \times 10^{-26}$ & $3.8 \times 10^{-38}$ \\
$e \mu$ & & $3.1 \times 10^{-7}$ & $5.3 \times 10^{-24}$ & $1.2 \times 10^{-21}$ & $1.4 \times 10^{-27}$ \\
$\mu \tau$ & $|C_{DL}^{\ell_1 \ell_2}/\Lambda^2| = $ & $2.6 \times 10^{-10}$ & $3.1 \times 10^{-28}$ & $1.2 \times 10^{-26}$ & FPS \\
$e \tau$ & & $2.7 \times 10^{-10}$ & $3.3 \times 10^{-28}$ & $1.3 \times 10^{-26}$ & $3.8 \times 10^{-38}$ \\
$e \mu$ & & $3.1 \times 10^{-7}$ & $5.3 \times 10^{-24}$ & $1.2 \times 10^{-21}$ & $1.4 \times 10^{-27}$ \\
\end{tabular}
\end{ruledtabular}
\end{table*}


The predicted upper limits of the four-fermion axial and pseudo-scalar operators for radiative pseudo-scalar decays $P \to \gamma \ell_1 \overline{\ell}_2$ in Tables \ref{tab:BdecaylimitsFF} and \ref{tab:DKdecaylimitsFF} demonstrate that these operators ultimately are better constrained by their two-body decay counterparts.  When we compare the predicted upper bounds of three-body rates in Tables \ref{tab:BdecaylimitsFF} and \ref{tab:DKdecaylimitsFF} to the two-body experimental limits in Table \ref{tab:Pdecaylimits} we see they are one to two orders of magnitude smaller.  Therefore the three-body decays could still provide complimentary access to these operators.


The tensor form factors in Appendix \ref{FormFactors} also allow us to analyze the contributions of the dipole operators of Eq. (\ref{eqn:LD}).
The dipole operators are best constrained via radiative lepton decays $\ell_2 \to \ell_1 \gamma$, where $\ell_2 = \tau \text{, } \mu$ and $\ell_1 = \mu \text{, } e$.  These decays have been the focus of most LFV experiments and therefore have the best constraints: ${\cal B}(\tau \to \mu \gamma) = 4.4 \times 10^{-8}$,  ${\cal B}(\tau \to e \gamma) = 3.3 \times 10^{-8}$, and ${\cal B}(\mu \to e \gamma) = 4.2 \times 10^{-13}$ \cite{PDG, Aubert:2009ag, TheMEG:2016wtm}.  In our previous work we were able to provide complimentary access via two-body vector quarkonium decays $V \to \gamma \ell_1 \overline{\ell}_2$ \cite{Hazard:2016fnc}.

Using the WC constraints obtained from the radiative lepton decays $\ell_2 \to \ell_1 \gamma$ in  \cite{Hazard:2016fnc}, we predict the dipole operator decay upper limits for $P \to \gamma \ell_1 \overline{\ell}_2$ in Table \ref{tab:DipoledecaylimitsFF}.  Here the predicted upper limits range from $10^{-21}$--$10^{-38}$, which is much lower than we would expect to be within experimental reach during the foreseeable future.  Despite showing that $P \to \gamma \ell_1 \overline \ell_2$ is not a useful means to constrain the dipole operators, the results in Table \ref{tab:DipoledecaylimitsFF} are ten or more orders of magnitude smaller than the predictions of the axial and pseudo-scalar operators in Tables \ref{tab:BdecaylimitsFF} and \ref{tab:DKdecaylimitsFF}.  This confirms that $P \to \gamma \ell_1 \overline{\ell}_2$ decays are better equiped to constrain four-fermion operators.  Indeed the operators in the best position to be constrained are the quark flavor changing four-fermion vector operators, which see no chiral suppression via lepton or quark masses and cannot be constrained via two-body decays.

\section{Conclusions}\label{Conclusions}

Studies of lepton flavor violating transitions are a promising path in the search for new physics. A convenient way to study new physics is to employ effective Lagrangians.
All models of new physics that include flavor-violating interactions 
are encoded in the values of Wilson coefficients of the low energy effective Lagrangian in Eq.~(\ref{eqn:Leff}). We argued that those Wilson coefficients 
can be constrained through the studies of radiative $B^0_q$, $\bar{D}^0$, and $K^0$ decays to two different flavored leptons. 

It is clear that studies of two-body $P \to \ell_1 \overline{\ell}_2$ decays allowed for the quantum number selection of a smaller subset of the effective operators, 
which reduced our reliance on single operator dominance.  Yet, the radiative three-body decays to $\gamma \ell_1 \overline{\ell}_2$ allowed access to the 
effective operators in Eq.~(\ref{eqn:Leff}) which cannot be probed via any two-body meson decays. In addition to probing new operators, the three-body 
radiative transitions also allowed for complimentary access to four-fermion operators constrained by two-body decays without the need to include a composite 
strongly-interacting meson to the final state. Finally, we provide evidence that the dipole operators are so well constrained by radiative LFV transitions 
$\ell_2 \to \ell_1 \gamma$ that their threshold for contributions to ${\cal B}(P \to \gamma \ell_1 \overline{\ell}_2)$ is many orders of magnitude below experimental reach. 
Thus, their contribution to the sum of amplitudes in Eq. (\ref{eqn:ScalarFunction}) can be safely dropped.

As more data is produced by Belle II and the LHCb experiment, we emphatically encourage our experimental colleagues to produce experimental limits on both 
LFV and radiative LFV decays of the $B^0_q$, $\bar{D}^0$, and $K^0$ mesons discussed in this work.

\begin{acknowledgments}
We would like to thank Dmitri Melikhov and Alexander Khodjamirian for useful discussions.
This work has been supported in part by the U.S. Department of Energy 
under contract DE-SC0007983. A.A.P. thanks the University of Siegen for hospitality where 
part of this work was performed under Comenius Guest Professorship. 
\end{acknowledgments}

\appendix
\section{Form Factors and Numerical Constants}\label{FormFactors}

To estimate differential decay rates and the upper limits of the total decay rates of the radiative decays in Section \ref{Results}, we must apply the form factors of 
Eqs. (\ref{eqn:axialvectorFF})--(\ref{eqn:offshelltensorFF}) and the numerical constants of Tables \ref{tab:MSbarMass} and \ref{tab:PengWC}.  Numerical inputs for the CKM matrix elements are found in \cite{PDG}.  
Before we can apply these form factors, we must relate them to those calculated in the literature, which are defined as 
\cite{Kruger:2002gf,Melikhov:2017pwu,Guadagnoli:2016erb,Kozachuk:2016ypz, Melikhov:2004mk}
\begin{eqnarray}
	\begin{split} \label{eqn:CalculatedFF}
		\langle \gamma^*(k_2)|\overline q_1 \gamma^\mu \gamma_5 q_2 | P(p) \rangle & =
			i e \varepsilon^*_{\alpha}(k_2) \left(g^{\alpha \mu} k_1 \cdot k_2 - k_1^{\alpha} k_2^{\mu} \right) \frac{F^P_A[k_1^2,k_2^2]}{m_P} \text{,} \\
		\langle \gamma^*(k_2)|\overline q_1 \gamma^\mu q_2 | P(p) \rangle & = \makebox[\widthof{$i$}][c]{}
			e \varepsilon^*_{\alpha}(k_2) \epsilon^{k_1 k_2 \mu \alpha} \frac{F^P_V[k_1^2,k_2^2]}{m_P} \text{,} \\
		\langle \gamma^*(k_2)|\overline q_1 \sigma^{\mu\nu} \gamma_5 q_2 | P(p) \rangle k_{1 \nu} & = \makebox[\widthof{$i$}][c]{}
			e \varepsilon^*_{\alpha}(k_2) \left(g^{\alpha \mu} k_1 \cdot k_2 - k_1^{\alpha} k_2^{\mu} \right) F^P_{TA}[k_1^2,k_2^2] \text{, and} \\
		\langle \gamma^*(k_2)|\overline q_1 \sigma^{\mu\nu} q_2 | P(p) \rangle k_{1\nu} & =
			i e \varepsilon^*_{\alpha}(k_2) \epsilon^{k_1 k_2 \mu \alpha} F^P_{TV}[k_1^2,k_2^2] \text{.} 
	\end{split}
\end{eqnarray}
\begin{table}
\caption{$\overline{MS}$ quark masses for decay calculations \cite{PDG}. \label{tab:MSbarMass}}
\begin{ruledtabular}
\begin{tabular}{ccccc} 
$m_u$& $m_d$ & $m_c$ & $m_s$ & $m_b$ \\
$2.2^{+0.6}_{-0.4}$ MeV & $4.7^{+0.5}_{-0.4}$ MeV  & $1.28 \pm 0.03$ GeV  & $96^{+8}_{-4}$ MeV  & $4.18^{+0.04}_{-0.03}$ GeV \\
\end{tabular}
\end{ruledtabular}
\end{table}

\begin{table}
\caption{Penguin operator Wilson coefficients, $C_{7 \gamma}$, for decay calculations. \label{tab:PengWC}}
\begin{ruledtabular}
\begin{tabular}{lccc} 
Transition & Scale $\mu$ [GeV] & $|C_{7 \gamma}|$ & Ref.\\
\hline
$b \to d(s) \gamma$ & $5.0$ & $0.299$ & \cite{Buchalla:1995vs} \\
$c \to u \gamma$ & $1.3$ & $\frac{0.0025}{4 \left| V_{ub}^{\ast} V_{cb} \right|}$ &\cite{Khodjamirian:2015dda} \\
\end{tabular}
\end{ruledtabular}
\end{table}

These form factors are functions of two momenta, $k_1$, which is emitted from the $q_1 \to q_2$ weak transition current, and $k_2$, which is emitted from one of the valence quarks of the meson $P$.  Here the photon is off-shell, but the on-shell definitions may be found by assuming $k_2^2=0$ and applying the momentum conservation relation $p = k_1 + k_2$.

Assuming $k^2 = 0$ and making the appropriate substitutions of $Q = p - k$ and $k$ for $k_1$ and $k_2$ we find the necessary relations between the form factors in Eqs. (\ref{eqn:axialvectorFF})--(\ref{eqn:offshelltensorFF}) and Eq. (\ref{eqn:CalculatedFF}) as

\begin{align} \label{eqn:FormFactorRelationships}
	\begin{split}
		F^P_{V \text{,}A}[Q^2,0] = \ & m_p f^P_{V \text{,}A}[Q^2,0] \text{,} \\
		F^P_{TV}[Q^2,0] = \ & -f^P_{T1}[Q^2,0] - p \cdot k f^P_{T2}[Q^2,0] \text{,} \\
		F^P_{TA}[Q^2,0] = \ & -f^P_{T1}[Q^2,0] - p \cdot Q f^P_{T2}[Q^2,0] \text{,} \\
		F^P_{TV, TA}[0,Q^2] = \ & - f^P_{T1}[0,Q^2] - f^P_{T3}[0,Q^2] \text{.}
	\end{split}
\end{align}

To make use of these relations we employ the parameterizations of  \cite{Kruger:2002gf} for the $F_V$, $F_A$, $F_{TV}$, and $F_{TA}$ form factors.  For the $B^0_q \to \gamma$ form factor parameterization when the photon $\gamma$ is emitted from the valence quarks ($k_1=Q$, $k_2=k$) we use 

\begin{equation} \label{eqn:Bformfactors}
F_i^{B_q}[E]=\beta_i \frac{f_P m_P}{\Delta_i+E_{\gamma}} \text{,} \quad i = \text{V, A, TV, TA}
\end{equation}

\noindent where $E_{\gamma}$ is the photon energy in the $P$-meson rest-frame. The constants $\beta$ and $\Delta$ are numerical parameters which can be found in Table \ref{tab:Bformfactors}.

\begin{table*}
\caption{ \label{tab:Bformfactors} Parameters of the $B^0_q \to \gamma$ form factors, as defined in Eq. (\ref{eqn:Bformfactors}) \cite{Kruger:2002gf}.}
\begin{ruledtabular}
\begin{tabular}{llcccc} & Parameter & $F_V$ & $F_{TV}$ & $F_A$ & $F_{TA}$ \\
\hline
$B^0_{d,s} \to \gamma$ & $\beta \left(\text{GeV}^{-1}\right)$ & 0.28 & 0.30 & 0.26 & 0.33 \\
& $\Delta \left(\text{GeV}\right)$ & 0.04 & 0.04 & 0.30 & 0.30 \\
\end{tabular}
\end{ruledtabular}
\end{table*}

For the parameterization of the $\bar{D}^0 \text{, } K^0 \to \gamma$ form factors when the photon $\gamma$ is emitted from the valence quarks ($k_1=Q$, $k_2=k$) we use 

\begin{equation} \label{eqn:DKformfactors}
F^{P}_{i}[m_{12}^2] = \frac{Q_{q_1} F_{i}^{(q_1)}[0] + Q_{q_2} F_{i}^{(q_2)}[0]}{1-\frac{m_{12}^2}{M_{i}^2}}\ \text{,} \quad i = \text{V, A, TV, TA} \text{.}
\end{equation}

\noindent Here $Q_{d(s)}= -\tfrac{1}{3}$, $Q_{u(c)}= \tfrac{2}{3}$, and the remaining parameters are found in Table \ref{tab:DKParametersFormFactors} \cite{Guadagnoli:2016erb}.

\begin{table*}
\caption{ \label{tab:DKParametersFormFactors} Parameters of the $\bar{D}^0 \text{,} K^0 \to \gamma$ form factors, as defined in Eq. (\ref{eqn:DKformfactors})  \cite{Guadagnoli:2016erb,Dmitri:PersonalCommunication}. The $K^0$ tensor form factors will be calculated elsewhere.}
\begin{ruledtabular}
\begin{tabular}{llcccc} & Parameter & $V$ & $A$ & $TV$ & $TA$ \\
\hline
$\bar{D}^0 \to \gamma$ & $F^c_i(0)$ & -0.12 & 0.14 & -0.12 & -0.12 \\
& $F^u_i(0)$ & -0.37 & -0.31 & -0.38 & -0.38 \\
& $M_{i}$ (GeV) & 2.0 & 2.3 & 2.0 & 2.4 \\
\hline
$K^0 \to \gamma$ & $F^d_i(0)$ & -0.22 & 0.20 & \textbf{--} &\textbf{--} \\
& $F^s_i(0)$ & -0.18 & -0.19 & \textbf{--} &\textbf{--} \\
& $M_{i}$ (GeV) & 0.89 & 0.89 & \textbf{--} &\textbf{--} \\
\end{tabular}
\end{ruledtabular}
\end{table*}

The form factors $F^P_{TV \text{, } TA}[0,Q^2]$ for $B^0_q$ and $\bar{D}^0$ decays are parameterized using vector meson dominance in  \cite{Melikhov:2004mk,Kozachuk:2016ypz}, which gives

\begin{equation}
F^P_{TV \text{, } TA}[0,Q^2] = F^P_{TV \text{, } TA}[0,0] - \sum_V 2 f_V g[0]_+^{P \to V} \frac{Q^2/m_V}{Q^2 - m_V^2 + i m_V \Gamma_V} \text{.}
\end{equation}

The vector meson dominance input parameter values are found in Table \ref{tab:VMDinputs}.  The $\rho$ and $\omega$ mesons are part of the vector meson sum for $B_d^0$ and $\bar{D}^0$ form factors because of their respective $d$ and $u$ valence quark content.  The $\phi$ meson is part of the vector meson sum for the $B^0_s$ form factor because of its $s$ valence quark content.  The zero momentum values of the tensor form factors are $F^{B_{d,s}^0}_{TV \text{, } TA}[0,0] = 0.115$ \cite{Kruger:2002gf} and $F^{\bar{D}^0}_{TV \text{, } TA}[0,0] = Q_c f^c_{TV,TA}[0] + Q_u f^u_{TV,TA}[0]$.

\begin{table} 
\caption{Vector meson dominance input parameters for $F_{TV \text{, } TA}[0,Q^2]$ form factors. \label{tab:VMDinputs}}
\begin{ruledtabular}
\begin{tabular}{l|ccccc|l}
$V$ & $g[0]_+^{B_q^0 \to V}$ & $g[0]_+^{\bar{D}^0 \to V}$ & $f_V$ (MeV) & $m_V$ (MeV) & $\Gamma_{V}$ (MeV) & Refs. \\
\hline
$\rho$ & $0.27$ & $-0.66$ & $154$ & $775.26 \pm 0.25$ & $147.8 \pm 0.9$ & \cite{PDG,Melikhov:2000yu,Melikhov:2004mk} \\
$\omega$ & $-0.27$ & $-0.66$ & $45.3$ & $782.65 \pm 0.12$ & $8.49 \pm 0.08$ & \cite{PDG,Melikhov:2000yu,Melikhov:2004mk} \\
$\phi$ & $-0.38$ & & $-58.8$ & $1019.460 \pm 0.016$ & $4.247 \pm 0.016$ &  \cite{PDG,Melikhov:2000yu,Melikhov:2004mk} \\
\end{tabular} 
\end{ruledtabular}
\end{table}

Given these form factors and the general input values given in Tables \ref{tab:MSbarMass} and \ref{tab:PengWC} we are able to plot the normalized differential decay rates and estimate the upper limits for the radiative branching ratios assuming single operator dominance in Section \ref{Results}.

\section{Quark Model}\label{QuarkModel}

When the necessary form factors are unavailable to take a model independent approach to the calculation of the four-fermion operator contributions of the diagrams in Fig. (\ref{3bodydecaydiagramsCD}), we may choose a model dependent approach.  We apply a constituent quark model to calculate the contributions of four-fermion vector, axial, and tensor operators of the type $(\ell_1 \overline{\ell}_2)( \overline{q} q)$.  We constrained both the vector and tensor Wilson coefficients for these operators previously in  \cite{Hazard:2016fnc}.  The results are reproduced here in Table \ref{tab:KnownWClimits} and can be used to find a predicted upper bound on the branching ratio of ${\cal B}\left(P \to \gamma \ell_1 \overline \ell_2 \right)$ for individual operators using the single operator dominance assumption.

\begin{table*}
\caption{Known Wilson coefficient limits from our previous work in  \cite{Hazard:2016fnc}. Note the center dots denote unknown values which could be constrained via $P \to \gamma \ell_1 \bar \ell_2$.  \label{tab:KnownWClimits}}
\begin{ruledtabular}
\begin{tabular}{cccccc}
& Leptons & \multicolumn{4}{c}{Quark} \\
Wilson coefficient $(\text{GeV}^{-2})$ & $ \overline{\ell_1} \ell_2$ & $b$ & $c$ & $s$ & $u/d$ \\
\hline
$|C_{VL(R)}^{q \ell_1 \ell_2}/\Lambda^2|$ & $\mu \tau$ & $3.5 \times 10^{-6}$ & $5.5 \times 10^{-5}$ & $\cdot \cdot \cdot$ & $\cdot \cdot \cdot$ \\
$|C_{VL(R)}^{q \ell_1 \ell_2}/\Lambda^2|$ & $e \tau$ & $4.1 \times 10^{-6}$ & $1.1 \times 10^{-4}$ & $\cdot \cdot \cdot$ & $\cdot \cdot \cdot$  \\
$|C_{VL(R)}^{q \ell_1 \ell_2}/\Lambda^2|$ & $e \mu$ & $\cdot \cdot \cdot$ & $1.0 \times 10^{-5}$ & $2.0 \times 10^{-3}$ & $\cdot \cdot \cdot$ \\
$|C_{AL(R)}^{q \ell_1 \ell_2}/\Lambda^2|$ & $e \mu$ & $\cdot \cdot \cdot$ & $\cdot \cdot \cdot$ & $2.0 \times 10^{-3}$ & $3.0 \times 10^{-3}$ \\
$|C_{TL(R)}^{q \ell_1 \ell_2}/\Lambda^2|$ & $\mu \tau$ & $2.8 \times 10^{-2}$ & $1.2$ & $\cdot \cdot \cdot$ & $\cdot \cdot \cdot$ \\
$|C_{TL(R)}^{q \ell_1 \ell_2}/\Lambda^2|$ & $e \tau$ & $3.2 \times 10^{-2}$ & $2.4$ & $\cdot \cdot \cdot$ & $\cdot \cdot \cdot$ \\
$|C_{TL(R)}^{q \ell_1 \ell_2}/\Lambda^2|$ & $e \mu$ & $\cdot \cdot \cdot$ & $4.8$ & $\cdot \cdot \cdot$ & $\cdot \cdot \cdot$ \\
\end{tabular}
\end{ruledtabular}
\end{table*}

\subsection{Consituent Quark Model} \label{Consituent Quark Model}

The amplitude for the diagrams in Fig. (\ref{3bodydecaydiagramsCD}) using this model is

\begin{align}
\begin{split}
i {\cal A}_{P \to \gamma \ell_1 \ell_2} = - \tfrac{i}{\Lambda^2} \varepsilon^{*\mu} \! (k) \sum_{i=1}^2 \Big( \overline{u}_{\ell_1} \! \left[\makebox[\widthof{$C_{TR}^{q_i \ell_1 \ell_2} \sigma^{\alpha \beta} P_L  + C_{TR}^{q_i \ell_1 \ell_2} \sigma^{\alpha \beta} P_R$}][c]{$C_{VR}^{q_i \ell_1 \ell_2} \gamma^{\alpha} P_R +  C_{VL}^{q_i \ell_1 \ell_2} \gamma^{\alpha} P_L$}\right] \!  v_{\ell_2} & \! \bra{0} \overline{q}_1 \Gamma^{V \text{,} q_i}_{\alpha \mu} q_2 \ket{P(p)}  \\
+ \overline{u}_{\ell_1} \! \left[\makebox[\widthof{$C_{TR}^{q_i \ell_1 \ell_2} \sigma^{\alpha \beta} P_L  + C_{TR}^{q_i \ell_1 \ell_2} \sigma^{\alpha \beta} P_R$}][c]{$C_{AR}^{q_i \ell_1 \ell_2} \gamma^{\alpha} P_R +  C_{AL}^{q_i \ell_1 \ell_2} \gamma^{\alpha} P_L $}\right] \! v_{\ell_2} & \! \bra{0} \overline{q}_1 \Gamma^{A \text{,} q_i}_{\alpha \mu} q_2 \ket{P(p)} \\
\quad \quad + m_2 m_{q_i} G_F \overline{u}_{\ell_1} \! \left[C_{TR}^{q_i \ell_1 \ell_2} \sigma^{\alpha \beta} P_L  + C_{TR}^{q_i \ell_1 \ell_2} \sigma^{\alpha \beta} P_R \right] \! v_{\ell_2} & \! \bra{0} \overline{q}_1 \Gamma^{T \text{,} q_i}_{\alpha \beta \mu} q_2 \ket{P(p)} \! \Big) \text{.}
\end{split}
\end{align}

This amplitude is dependent on matrix elements of the form $\bra{0} \overline{q}_1 \Gamma q_2 \ket{P}$ with the matrices $\Gamma$ defined for each operator (${\cal O} \sim (\ell_1 \overline{\ell}_2) (\overline{q}_i q_i)$, $i = 1,2$) as 

\begin{align}
	\begin{split}  \label{eqn:q1q1l1l2ME}
		\Gamma^{V \text{,} q_1}_{\alpha \mu} = \ & i \tfrac{G_F}{\sqrt{2}} \tfrac{\sqrt{4 \pi \alpha}}{\pi^2} m_{q_1} C_{7 \gamma} \sum_q \lambda_q^P \gamma_{\alpha}
			\tfrac{x\slashed{p} - \slashed{k} + m_{q_1}}{\left( x p - k\right)^2 - m_{q_1}^2} \sigma_{\mu \nu} \left(1 + \gamma_{5}\right) k^{\nu} \text{,} \\
		\Gamma^{A \text{,} q_1}_{\alpha \mu} = \ & i \tfrac{G_F}{\sqrt{2}} \tfrac{\sqrt{4 \pi \alpha}}{\pi^2} m_{q_1} C_{7 \gamma} \sum_q \lambda_q^P \gamma_{\alpha}
			\gamma_5 \tfrac{x\slashed{p} - \slashed{k} + m_{q_1}}{\left( x p - k\right)^2 - m_{q_1}^2} \sigma_{\mu \nu} \left(1 + \gamma_{5}\right) k^{\nu} \text{,} \\
		\Gamma^{T  \text{,} q_1}_{\alpha \beta \mu} = \ & i \tfrac{G_F}{\sqrt{2}} \tfrac{\sqrt{4 \pi \alpha}}{\pi^2} m_{q_1} C_{7 \gamma} \sum_q \lambda_q^P 
			\sigma_{\alpha \beta} \tfrac{x\slashed{p} - \slashed{k} + m_{q_1}}{\left( x p - k\right)^2 - m_{q_1}^2} \sigma_{\mu \nu} 
			\left(1 + \gamma_{5}\right) k^{\nu} \text{,} 
	\end{split}
\end{align}

\begin{align}
	\begin{split}  \label{eqn:q2q2l1l2ME}
		\Gamma^{V \text{,} q_2}_{\alpha \mu} = \ & i \tfrac{G_F}{\sqrt{2}} \tfrac{\sqrt{4 \pi \alpha}}{\pi^2} m_{q_2} C_{7 \gamma} \sum_q \lambda_q^P \sigma_{\mu \nu}
			\left(1 + \gamma_{5}\right) k^{\nu} \tfrac{-\left(1-x\right) \slashed{p} + \slashed{k} + m_{q_2}}{\left( \left(1-x\right)p - k\right)^2 - m_{q_2}^2}
			\gamma_{\alpha} \text{,} \\
		\Gamma^{A \text{,} q_2}_{\alpha \mu} = \ & i \tfrac{G_F}{\sqrt{2}} \tfrac{\sqrt{4 \pi \alpha}}{\pi^2} m_{q_2} C_{7 \gamma} \sum_q \lambda_q^P \sigma_{\mu \nu}
		\left(1 + \gamma_{5}\right) k^{\nu} \tfrac{-\left(1-x\right) \slashed{p} + \slashed{k} + m_{q_2}}{\left( \left(1-x\right)p - k\right)^2 - m_{q_2}^2} \gamma_{\alpha}
		\gamma_5 \text{, and} \\
		\Gamma^{T  \text{,} q_2}_{\alpha \beta \mu} = \ & i \tfrac{G_F}{\sqrt{2}} \tfrac{\sqrt{4 \pi \alpha}}{\pi^2} m_{q_2} C_{7 \gamma} \sum_q \lambda_q^P 
			\sigma_{\mu \nu} \left(1 + \gamma_{5}\right)  k^{\nu} \tfrac{-\left(1-x\right) \slashed{p} + \slashed{k} + m_{q_2}}
			{\left( \left(1-x\right)p - k\right)^2 - m_{q_2}^2} \sigma_{\alpha \beta} \text{.}
	\end{split}
\end{align}


In modeling the quark anti-quark distribution, we chose to follow  \cite{Aditya:2012ay,Szczepaniak:1990dt, Lepage:1980fj}, where we can write the wave function of the ground state, $P(p)$, as

\begin{equation}\label{QMwavefunc}
\psi_P = \frac{I_c}{\sqrt{6}} \phi_P[x] \gamma_5 \left(\slashed{p} + m_P g[x] \right).
\end{equation}

The variable $x$ is the momentum fraction of one of the quarks and $I_c$ is the identity matrix of color space.  We have assigned the momenta in Fig. (\ref{3bodydecaydiagramsCD}) such that the valence quark $\overline{q}_1$ has momentum $x P$ and the valence quark $q_2$ has momentum $(1-x) P$.  The function $g_P[x]$ is $g_H[x] \sim 1$ for heavy mesons and $g_L[x]=0$ for light mesons.  The distribution amplitudes used for light and heavy mesons and their normalization are 

\begin{align}\label{QMdistributionamp}
\begin{split}
\phi_L \sim \ & x\left(1-x\right) \text{,}\\
\phi_H \sim \ & \left[\frac{m_{q_L}}{M_H}\frac{1}{1-x}+\frac{1}{x}-1\right]^{-2} \text{,}\\
\frac{f_P}{2 \sqrt{6}} = \ & \int_0^1 \phi[x]~dx  \text{.}
\end{split}
\end{align}

Here $m_{q_L}$ is the mass of the light quark and the normalization is related to the decay constant $f_P$.  By taking the trace and integrating over the momentum fraction we find the matrix element

\begin{equation}\label{QMmatrixelement}
\bra{0} \overline{q}_1 \Gamma^{\mu} q_2 \ket{P} = \int_0^1 \text{Tr}[\Gamma^{\mu} \psi_P]~dx  \text{.}
\end{equation}

\subsection{Spectra and Limits} \label{ResultsQM}

\begin{table*} 
\caption{Constituent quark masses used in calculations of quark model matrix element \cite{Scadron:2006dy}. \label{tab:ConstituentQuarkMasses}}
\begin{ruledtabular}
\begin{tabular}{|cccccc|}
Quark & $m_u$ & $m_d$ & $m_s$ & $m_c$ & $m_b$ \\
\hline
Constituent mass (MeV) & $335.5$ & $339.5$ & $486$ & $1550$ & $4730$ \\
\end{tabular}
\end{ruledtabular}
\end{table*}

\begin{figure}
\subfigure[]{\includegraphics[scale=0.41]{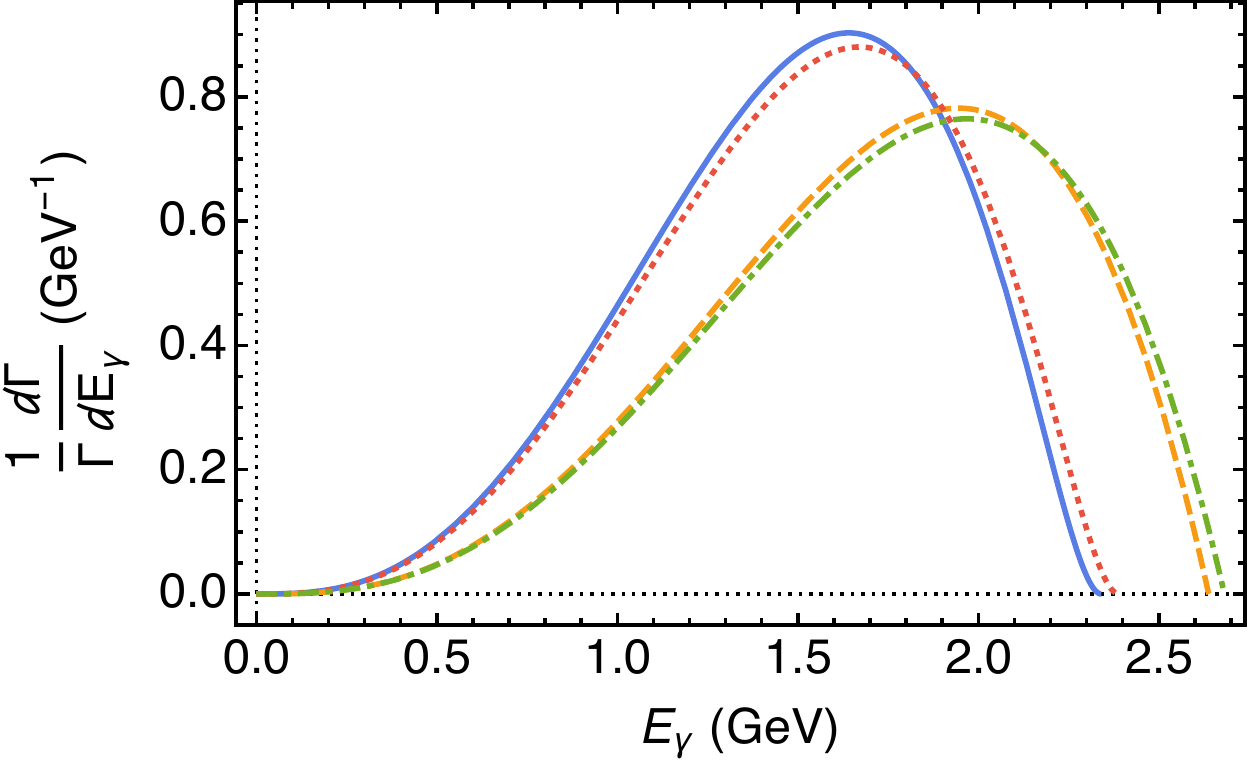} \label{fig:bl1l2vectordecayplotsQM}}
\subfigure[]{\includegraphics[scale=0.41]{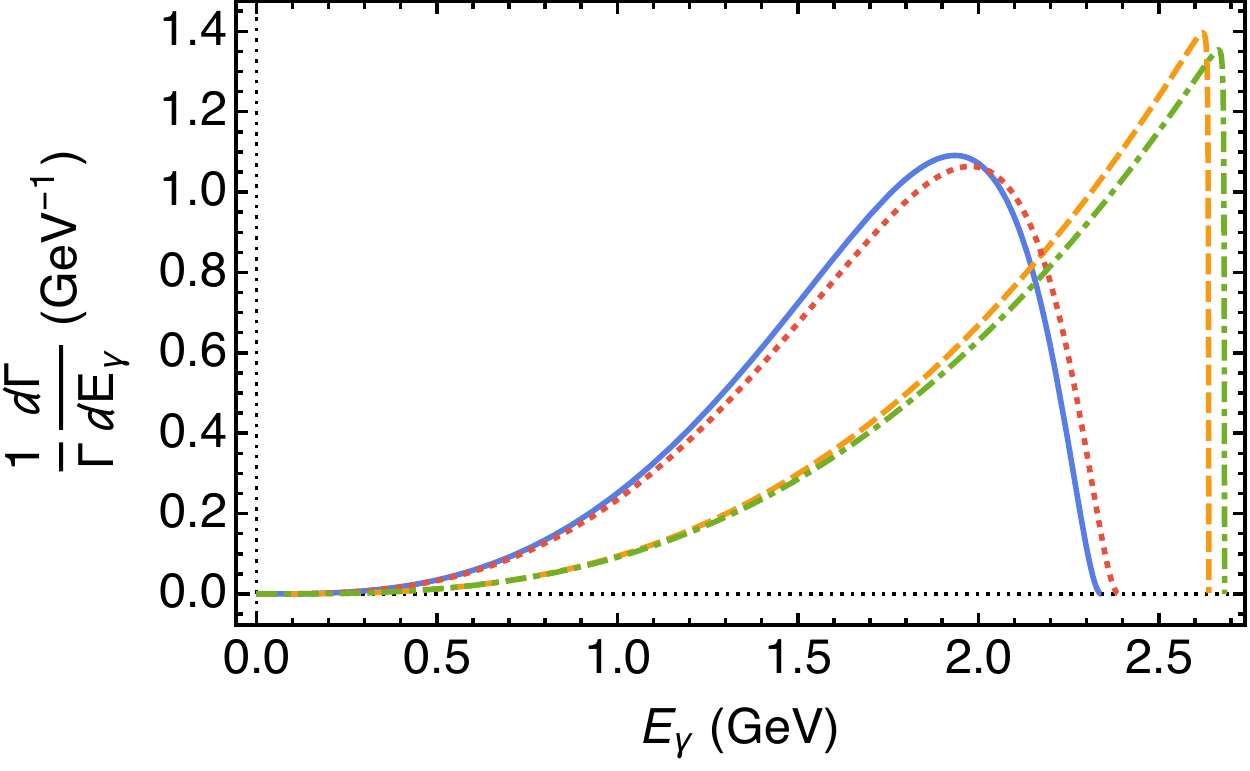} \label{fig:bl1l2tensorLdecayplotsQM}}
\subfigure[]{\includegraphics[scale=0.41]{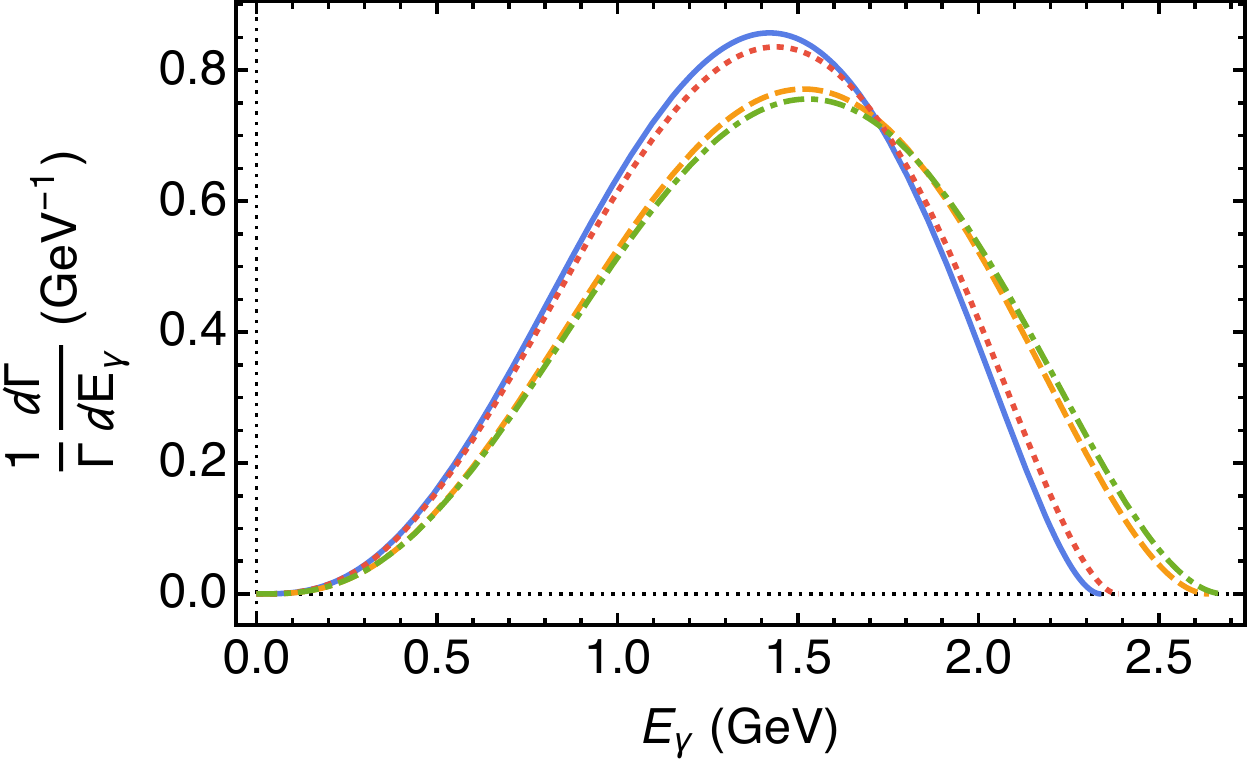} \label{fig:bl1l2tensorRdecayplotsQM}}
\caption{Differential decay plots as functions of photon energy $E_{\gamma}$ for (a) vector/axial, (b) left-handed tensor, and (c) right-handed tensor operators of the type ${\cal O} \sim \left(\ell_1 \overline{\ell}_2\right)\left(\overline{b}b\right)$. Plotted decay rates are $B_d \to \gamma \mu \tau$ or $\gamma e \tau$ (solid blue curve), $B_d \to \gamma e \mu$ (short-dashed gold curve), $B_s \to \gamma \mu \tau$ or $\gamma e \tau$ (dotted red curve), $B_s \to \gamma e \mu$ (dot-dashed green curve).\label{fig:bl1l2decayplotsQM}}
\end{figure}

Since we applied a constituent quark model to calculate the transition amplitudes we need to define its parameters (constituent quark mass) that are used to calculate the matrix element in Eq. (\ref{QMmatrixelement}).  These masses are in Table \ref{tab:ConstituentQuarkMasses}. Using this matrix element and integrating over the Mandelstam variables $m_{23}^2$ and $m_{12}^2$ we can calculate the differential decay rate as a function of the photon energy, $E_{\gamma}$, in the rest-frame of the meson $P$ and the total decay rate.  An example plot for these differential decay spectra normalized to the total decay rate is Fig. (\ref{fig:bl1l2decayplotsQM}), which shows the spectra of $B^0_q$ decays for the vector, axial, and tensor operators of type $(\ell_1 \overline{\ell}_2)( \overline{q} q)$. The normalization cancels out sources of uncertainty such as the Wilson coefficients (i.e. $C_{VR(L)}^{q_i \ell_1 \ell_2}$) and the CKM matrix element values.  As we did in Section \ref{Limits}, we 
apply known Wilson coefficient constraints from Table. \ref{tab:KnownWClimits} and the single operator dominance assumption to the total decay rate to make predictions of the branching ratio upper limit for these operators, which can be found in Tables. \ref{tab:BdecaylimitsQM} and \ref{tab:DdecaylimitsQM}.

\begin{table*}
\caption{\label{tab:BdecaylimitsQM} Upper limits on $B_{q}^0(q \bar b) \to \gamma \ell_1 \overline{\ell}_2$ branching ratios from known Wilson coefficient constraints using constituent quark model.  The center dots indicate no Wilson coefficient constraints were available for a prediction of an upper bound.  Experimental studies of this decay channel would present an opportunity to constrain these Wilson coefficients. }
\begin{ruledtabular}
\begin{tabular}{lcccccc} Wilson & \multicolumn{6}{c}{Upper limits} \\
coefficient & ${\cal B}(B^0_d \to \gamma \mu \tau)$ & ${\cal B}(B^0_d \to \gamma e \tau)$ & ${\cal B}(B^0_d \to \gamma e \mu)$ & ${\cal B}(B^0_s \to \gamma \mu \tau)$ &${\cal B}(B^0_s \to \gamma e \tau)$ & ${\cal B}(B^0_s \to \gamma e \mu)$ \\
\hline
$C_{VR}^{b \ell_1 \ell_2}$ & $5.7 \times 10^{-20}$ & $7.8 \times 10^{-20}$ & $\cdot \cdot \cdot$ & $1.8 \times 10^{-18}$ & $2.5 \times 10^{-18}$ & $\cdot \cdot \cdot$ \\
$C_{VL}^{b \ell_1 \ell_2}$ & $5.7 \times 10^{-20}$ & $7.8 \times 10^{-20}$ & $\cdot \cdot \cdot$ & $1.8 \times 10^{-18}$ & $2.5 \times 10^{-18}$ & $\cdot \cdot \cdot$ \\
$C_{VR}^{q \ell_1 \ell_2}$ & $\cdot \cdot \cdot$ & $\cdot \cdot \cdot$ & $\cdot \cdot \cdot$ & $\cdot \cdot \cdot$ & $\cdot \cdot \cdot$ & $1.3 \times 10^{-10}$ \\
$C_{VL}^{q \ell_1 \ell_2}$ & $\cdot \cdot \cdot$ & $\cdot \cdot \cdot$ & $\cdot \cdot \cdot$ & $\cdot \cdot \cdot$ & $\cdot \cdot \cdot$ & $1.3 \times 10^{-10}$ \\
$C_{AR}^{q \ell_1 \ell_2}$ & $\cdot \cdot \cdot$ & $\cdot \cdot \cdot$ & $2.0 \times 10^{-12}$ & $\cdot \cdot \cdot$ & $\cdot \cdot \cdot$ & $1.5 \times 10^{-11}$ \\
$C_{AL}^{q \ell_1 \ell_2}$ & $\cdot \cdot \cdot$ & $\cdot \cdot \cdot$ & $2.0 \times 10^{-12}$ & $\cdot \cdot \cdot$ & $\cdot \cdot \cdot$ & $1.5 \times 10^{-11}$ \\
$C_{TR}^{b \ell_1 \ell_2}$ & $3.9 \times 10^{-21}$ & $5.1 \times 10^{-21}$ & $\cdot \cdot \cdot$ & $2.1 \times 10^{-19}$ & $2.8 \times 10^{-19}$ & $\cdot \cdot \cdot$ \\
$C_{TL}^{b \ell_1 \ell_2}$ & $1.1 \times 10^{-18}$ & $1.5 \times 10^{-18}$ & $\cdot \cdot \cdot$ & $3.9 \times 10^{-17}$ & $5.1 \times 10^{-17}$ & $\cdot \cdot \cdot$ \\
\end{tabular}
\end{ruledtabular}
\end{table*}

These limits range in order of magnitude from $10^{-10}$--$10^{-28}$ and therefore many are below experimental reach.  It is the spaces between these limits that should draw the reader's attention.  There is much opportunity here to constrain the operators whose limits cannot be predicted.  Providing limits using these RLFV decays would of course be complementary to two-body LFV decays of quarkonia (e.g. \cite{Hazard:2016fnc}), but would come for free as we constrain the vector and tensor operators with flavor changes on both the quark and lepton sides.

\begin{table*}
\caption{\label{tab:DdecaylimitsQM} Upper limits on $\bar{D}^0 \left(u \bar c \right) \to \gamma \ell_1 \overline{\ell}_2$ branching ratios from known Wilson coefficient constraints using constituent quark model.  The center dots indicate no Wilson coefficient constraints were available for a prediction of an upper bound.  Experimental studies of this decay channel would present an opportunity to constrain these Wilson coefficients.}
\begin{ruledtabular}
\begin{tabular}{lcc} Wilson & \multicolumn{2}{c}{Upper limits} \\
coefficient & ${\cal B}(\bar{D}^0 \to \gamma e \tau)$ & ${\cal B}(\bar{D}^0 \to \gamma e \mu)$ \\
\hline
$C_{VR}^{c \ell_1 \ell_2}$ & $5.1 \times 10^{-28}$ & $8.8 \times 10^{-24}$ \\
$C_{VL}^{c \ell_1 \ell_2}$ & $5.1 \times 10^{-28}$ & $8.8 \times 10^{-24}$ \\
$C_{AR}^{u \ell_1 \ell_2}$ & $\cdot \cdot \cdot$ & $1.3 \times 10^{-16}$ \\
$C_{AL}^{u \ell_1 \ell_2}$ & $\cdot \cdot \cdot$ & $1.3 \times 10^{-16}$ \\
$C_{TR}^{c \ell_1 \ell_2}$ & $6.0 \times 10^{-28}$ & $2.5 \times 10^{-24}$ \\
$C_{TL}^{c \ell_1 \ell_2}$ & $6.2 \times 10^{-27}$ & $3.7 \times 10^{-22}$ \\
\end{tabular}
\end{ruledtabular}
\end{table*}




\end{document}